\documentclass[aip]{revtex4}

\usepackage{amsmath}
\usepackage{amssymb}
\usepackage{stmaryrd}

\usepackage{graphicx}

\usepackage{epsfig}

\def\ov#1{\overline{#1}}

\def\vb#1{\mbox{\boldmath$#1$}}
\def\pd#1#2{\frac{\partial #1}{\partial #2}}

\def\wh#1{\widehat{#1}}
\def\bdot{\,\vb{\cdot}\,}
\def\btimes{\,\vb{\times}\,}

\def\cal#1{\mathcal{#1}}

\newcommand{\bc}{\begin{center}}
\newcommand{\ec}{\end{center}}
\newcommand{\bt}{\begin{tabbing}}
\newcommand{\et}{\end{tabbing}}
\newcommand{\be}{\begin{eqnarray*}}
\newcommand{\ee}{\end{eqnarray*}}
\newcommand{\bs}{\begin{slide}}
\newcommand{\es}{\end{slide}}

\begin{document}

\title{Scenarios for magnetic X-point collapse in 2D incompressible \\ dissipationless extended magnetohydrodynamics}

\author{Alain J.~Brizard}
\affiliation{Department of Physics, Saint Michael's College, Colchester, VT 05439, USA}

\begin{abstract}
The equations of 2D incompressible dissipationless extended magnetohydrodynamics (XMHD) extend the equations of incompressible Hall MHD (HMHD) by retaining finite-electron inertia. These XMHD equations couple the fluid velocity ${\bf V} = \wh{\sf z}\btimes\nabla\phi + V_{z}\,\wh{\sf z}$ with the magnetic field ${\bf B} = \nabla\psi\btimes\wh{\sf z} + B_{z}\,\wh{\sf z}$ in a process that is known to support dissipationless solutions that exhibit finite-time singularities associated with magnetic X-point collapse in the magnetic plane $(B_{x} = \partial\psi/\partial y, B_{y} = -\,\partial\psi/\partial x)$. Here, by adopting a 2D self-similar model for the four XMHD fields $(\phi,\psi,V_{z},B_{z})$, we obtain five coupled ordinary differential equations that are solved in terms of the Jacobi elliptic functions based on an orbital classification associated with particle motion in a quartic potential. Excellent agreement is found when these analytical solutions are compared with numerical solutions, including the precise time of a magnetic X-point collapse.
\end{abstract}

\date{\today}

\maketitle

\section{Introduction}

The problem of the dissipationless collapse of a magnetic X-point \cite{PTR_1995,Priest_2016,ZY_2016,Priest_2022} is a paradigm problem in plasma physics that requires an extended formulation beyond ideal magnetohydrodynamics (MHD) \cite{Charidakos_2014,Abdelhamid_2015}, where two-fluid effects (e.g., the Hall term and finite electron inertia) are retained. These dissipationless extended MHD (XMHD) equations \cite{Kimura_2014,Lingham_2015a,Lingham_2015b,Grasso_2017}  are parametrized by the plasma dimensionless parameters ${\sf d}_{i} = c/(\omega_{\rm pi}L)$ and ${\sf d}_{e} \equiv \sqrt{\delta}\,{\sf d}_{i}$, which denote the ion and electron skin depths (with a finite electron-ion mass ratio $\delta \equiv m_{e}/m_{i} \ll 1$) normalized to a characteristic length scale $L$. We note, on the one hand, that the dissipationless XMHD equations include the dissipationless Hall MHD (HMHD) equations \cite{Huba_1995} (${\sf d}_{i} \neq 0$ and ${\sf d}_{e} = 0$) and the dissipationless inertial MHD equations \cite{Kimura_2014} (${\sf d}_{i} = 0$ and ${\sf d}_{e} \neq 0$), and these dissipationless reduced MHD equations possess solutions that exhibit magnetic X-point collapse singularities that occur in a finite time \cite{Ottaviani_1993,Nunez_2004,Dreher_2005,Litvinenko_2007,Litvinenko_2015a,Litvinenko_2015b,Janda_2018,Brizard_2019,Janda_2019}. We note that the equations of ideal MHD \cite{Freidberg_1982,Taylor_1986} (${\sf d}_{i} = 0 = {\sf d}_{e}$), on the other hand, cannot break the magnetic topology needed for a dissipationless magnetic X-point collapse.

\subsection{Incompressible dissipationless extended MHD equations}

The extended MHD study of the dissipationless collapse of a magnetic X-point is based on the following coupled incompressible fluid equations (normalized to standard Alfv\'{e}nic units) \cite{Grasso_2017}: the fluid momentum equation
\begin{equation}
\pd{\bf V}{t} + {\bf V}\bdot\nabla{\bf V} \;=\; -\,\nabla P \;+\; {\bf J}\btimes{\bf B} \;-\; {\sf d}_{e}^{2}\;{\bf J}\bdot\nabla{\bf J}, \label{eq:V_fluid}
\end{equation}
where ${\bf V}$ denotes the normalized center-of-mass fluid velocity, the normalized fluid mass density $\varrho = 1$ has been set to unity (so that the fluid continuity equation yields the incompressibility condition $\nabla\bdot{\bf V} \equiv 0$), $P$ denotes the normalized total fluid pressure, and ${\bf J} \equiv \nabla\btimes{\bf B}$ denotes the normalized Amp\`{e}re current density obtained from the normalized magnetic field ${\bf B} \equiv \nabla\btimes{\bf A}$; the generalized Ohm's law
\begin{equation} 
{\bf E} + {\bf V}\btimes{\bf B} \;=\; {\sf d}_{i}\left({\bf J}\btimes{\bf B} - \nabla P_{e}\right) \;+\; {\sf d}_{e}^{2}\left(\pd{\bf J}{t} + {\bf V}\bdot\nabla{\bf J} + {\bf J}\bdot\nabla{\bf V}\right) \;-\; {\sf d}_{i}{\sf d}_{e}^{2}\,{\bf J}\bdot\nabla{\bf J}, \label{eq:Ohm_general} 
\end{equation}
where ${\bf E} \equiv -\,\nabla\phi - \partial{\bf A}/\partial t$ denotes the normalized electric field and $P_{e}$ denotes the normalized electron pressure; and Faraday's law
\begin{equation}
\pd{\bf B}{t} \;=\; -\;\nabla\btimes{\bf E}, \label{eq:Faraday}
\end{equation}
which is subject to the condition $\nabla\bdot{\bf B} \equiv 0$. Here, terms that involve ${\sf d}_{i} \neq 0$ represent Hall corrections while terms that involve ${\sf d}_{e} \equiv \sqrt{\delta}\,{\sf d}_{i} \neq 0$ represent finite electron-inertia corrections. 

While the nature of three-dimensional (3D) magnetic reconnection can be very different from two-dimensional (2D) magnetic reconnection \cite{Priest_2016,Priest_2022}, we focus our attention on 2D dissipationless magnetic reconnection \cite{Porcelli_2002,Borgogno_2005} in the present paper because of its simplicity and its relevance in many physical situations. Here, we assume that the fields are independent of the $z$-coordinate, so that the incompressible fluid velocity ${\bf V} = \wh{\sf z}\btimes\nabla\phi + V_{z}\,\wh{\sf z}$ and the magnetic field ${\bf B} = \nabla\psi\btimes\wh{\sf z} + B_{z}\,\wh{\sf z}$ are now defined in terms of the 2D four fields $(\phi,V_{z},\psi,B_{z})$, where ${\bf A} \equiv \psi\,\wh{\sf z} + {\bf A}_{\bot}$, $B_{z} \equiv \wh{\sf z}\bdot\nabla\btimes{\bf A}_{\bot}$, and ${\bf J} = -\,\nabla^{2}\psi\,\wh{\sf z} + \nabla B_{z}\btimes\wh{\sf z}$. The evolution equations for the four XMHD fields $(\phi,V_{z},\psi,B_{z})$ are extracted from Eqs.~\eqref{eq:V_fluid}-\eqref{eq:Faraday} as follows \cite{Grasso_2017}. First, we take the $z$-component of the momentum equation \eqref{eq:V_fluid} to obtain
\begin{equation}
\frac{dV_{z}}{dt} \;\equiv\; \pd{V_{z}}{t} + \left[\phi,\;  V_{z}\right] \;=\; \left[B_{z},\; \psi^{*}\right], \label{eq:Vz_dot}
\end{equation}
where the antisymmetric spatial bracket $[f,\; g] \equiv \wh{\sf z}\bdot\nabla f\btimes\nabla g$ is expressed in terms of normalized spatial coordinates $(x,y)$, and $\psi^{*} \equiv \psi - {\sf d}_{e}^{2}\nabla^{2}\psi$ includes a finite electron-inertia correction \cite{Kimura_2014,Lingham_2015a,Lingham_2015b,Grasso_2017}. Second, we take the curl of the momentum equation \eqref{eq:V_fluid}, which yields the vorticity equation for $\wh{\sf z}\bdot\nabla\btimes{\bf V} = \nabla^{2}\phi$:
\begin{equation}
\frac{d\nabla^{2}\phi}{dt} \;\equiv\; \pd{\nabla^{2}\phi}{t} + \left[\phi,\;  \nabla^{2}\phi\right] \;=\; \left[\psi,\;  \nabla^{2}\psi\right] \;-\; {\sf d}_{e}^{2}\,\left[B_{z},\;  \nabla^{2}B_{z}\right], \label{eq:omega_dot}
\end{equation}
Third, we take the $z$-component of the generalized Ohm's Law \eqref{eq:Ohm_general}, which yields
\begin{equation}
\frac{d\psi^{*}}{dt} \;\equiv\; \pd{\psi^{*}}{t} + \left[\phi,\;  \psi^{*}\right] \;=\; -\,{\sf d}_{i}\,\left[B_{z},\;  \psi^{*}\right] \;+\; {\sf d}_{e}^{2}\,\left[B_{z},\;  V_{z}\right], \label{eq:psi_dot}
\end{equation}
while, fourth, the $z$-component of Faraday's Law \eqref{eq:Faraday} yields
\begin{equation}
\frac{dB_{z}^{*}}{dt} \;\equiv\; \pd{B_{z}^{*}}{t} + \left[\phi,\;  B_{z}^{*}\right] \;=\; {\sf d}_{i}\,\left[\nabla^{2}\psi,\;  \psi\right] \;+\; \left[V_{z},\;  \psi\right] \;+\; {\sf d}_{e}^{2}\,\left(\left[B_{z},\;  \nabla^{2}\phi\right] \;+\frac{}{} {\sf d}_{i}\,\left[B_{z},\;  \nabla^{2}B_{z}\right] \right), \label{eq:Bz_dot}
\end{equation}
where $B_{z}^{*} \equiv B_{z} - {\sf d}_{e}^{2}\nabla^{2}B_{z}$ includes finite electron-inertia correction  \cite{Kimura_2014,Lingham_2015a,Lingham_2015b,Grasso_2017}. We note that Eqs.~\eqref{eq:Vz_dot}-\eqref{eq:Bz_dot} may be combined in terms of the two coupled equations
\begin{eqnarray}
\frac{d}{dt}\left({\sf d}_{i}V_{z} \;+\frac{}{} \psi^{*}\right) &=& {\sf d}_{e}^{2}\,[B_{z},\; V_{z}], \label{eq:V_psi} \\
\frac{d}{dt}\left(B_{z}^{*} \;+\frac{}{} {\sf d}_{i}\,\nabla^{2}\phi\right) &=& [V_{z},\;\psi] \;+\;  {\sf d}_{e}^{2}\,[B_{z},\; \nabla^{2}\phi], \label{B_phi}
\end{eqnarray}
which consider the time evolutions of the $z$-component of the canonical momentum $\wh{\sf z}\bdot{\bf P}^{*} \;=\; \wh{\sf z}\bdot\left({\sf d}_{i}\,{\bf V} + {\bf A}^{*}\right) = {\sf d}_{i}V_{z} + \psi^{*}$ and its parallel vorticity $\wh{\sf z}\bdot\nabla\btimes{\bf P}^{*} \;=\; 
{\sf d}_{i}\,\nabla^{2}\phi + B_{z}^{*}$. The Hamiltonian properties of the 2D XMHD equations \eqref{eq:Vz_dot}-\eqref{eq:Bz_dot}, including the associated Casimir invariants, have been studied extensively elsewhere \cite{Kimura_2014,Abdelhamid_2015}. Here, we point out that, according to dissipationless Hall MHD (${\sf d}_{e} = 0$), we obtain the conservation law 
\begin{equation}
\frac{d}{dt}\left({\sf d}_{i}\,V_{z} \;+\frac{}{} \psi\right) \;=\; 0 
\label{eq:Hall_psi}
\end{equation}
from Eq.~\eqref{eq:V_psi}, which breaks the ideal MHD (${\sf d}_{i} = 0$) frozen-flux condition $\wh{\sf z}\bdot({\bf E} + {\bf V}\btimes{\bf B}) = -\,d\psi/dt = 0$ and allows a change in magnetic topology in the $(x,y)$-plane.

Lastly, we note that, without the axial fields $V_{z} = 0 = B_{z}$, which bypasses the Hall term in Eq.~\eqref{eq:psi_dot}, the two-field dissipationless equations $d\nabla^{2}\phi/dt = [\psi,\;  \nabla^{2}\psi]$ and $d\psi^{*}/dt = 0$ can also be used to study collisionless magnetic reconnection \cite{Ottaviani_1993,Ottaviani_1995}. Other dissipationless two-field extended MHD models used to study dissipationless 2D magnetic reconnection (where the magnetic field ${\bf B} = B_{0}\,\wh{\sf z} + \nabla\psi\btimes\wh{z}$ includes a constant guide field $B_{0}$) have appeared in the literature \cite{Cafaro_1998,Grasso_1999,Tassi_2010}, where a new dimensionless parameter ${\sf d}_{s} = (c_{s}/\omega_{ci})/L \equiv \sqrt{\beta_{e}}\,{\sf d}_{i}$, which is defined as the normalized ratio of the ion sound speed $c_{s}$ to the ion cyclotron frequency $\omega_{ci}$, is introduced (associated with the inclusion of parallel electron compressibility, with $\beta_{e} = 4\pi\,n_{0}T_{e}/B_{0}^{2}$), so that we obtain a new set of two-field equations become $d\nabla^{2}\phi/dt = [\psi,\;  \nabla^{2}\psi]$ and $d\psi^{*}/dt = {\sf d}_{s}^{2}\,[\nabla^{2}\phi,\; \psi]$. We note that these equations can be combined into the coupled equations $\partial\Psi^{\pm}/\partial t + [\Phi^{\pm},\; \Psi^{\pm}] = 0$, where $\Psi^{\pm} \equiv \psi^{*} \pm {\sf d}_{e}{\sf d}_{s}\,\nabla^{2}\phi$ and $\Phi^{\pm} \equiv \phi \pm ({\sf d}_{s}/{\sf d}_{e})\,\psi$. While these dissipationless MHD equations can be generalized to a four-field model \cite{Tassi_2010}, once again used to investigate fast dissipationless magnetic reconnection, we will restrict our attention to the standard XMHD equations \eqref{eq:Vz_dot}-\eqref{eq:Bz_dot} in what follows.

\subsection{Standard 2D self-similar XMHD model}

Since the detailed solution of the 2D nonlinear XMHD equations \eqref{eq:Vz_dot}-\eqref{eq:Bz_dot} requires extensive numerical simulations, it would be advantageous to proceed with a simple representation of the XMHD fields  $(\phi,V_{z},\psi,B_{z})$ in order to extract the key elements of finite-time singularities associated with dissipationless X-point magnetic collapse. For this purpose, a standard 2D self-similar XMHD model \cite{Litvinenko_2007,Litvinenko_2015a,Litvinenko_2015b,Janda_2018,Brizard_2019,Janda_2019}  is introduced through the scalar fields
\begin{equation}
\left. \begin{array}{rcl}
\phi(x,y,t) &=& \gamma(t)\,xy \\
{\sf d}_{i}\,V_{z}(x,y,t) &=&  u(t)\,x^{2} \;+\; v(t)\,y^{2} \\
\psi^{*}(x,y,t) &=& \imath(t)\,x^{2} \;-\; \jmath(t)\,y^{2} \\
{\sf d}_{i}\,B_{z}(x,y,t) &=& b(t)\,x y
\end{array} \right\},
\label{eq:XMHD_coef}
\end{equation}
where the five coefficients $(\imath,\jmath,u,v,b)$ are arbitrary functions of time. Here, Eq.~\eqref{eq:omega_dot} is trivially satisfied, since $\nabla^{2}\phi = 0 = \nabla^{2}B_{z}$, while $\psi = \imath\,x^{2} - \jmath\,y^{2} + 2\,{\sf d}_{e}^{2}\,(\imath - \jmath)$ yields
a uniform parallel current density $J_{z} = -\,\nabla^{2}\psi = 2\,(\jmath - \imath)$ so that $\nabla(\nabla^{2}\psi) = 0$. We note, however, that the 2D self-similar XMHD model \eqref{eq:XMHD_coef} is not unique in studying magnetic reconnection and other models \cite{Craig_2003a,Craig_2003b,Craig_2005,Litvinenko_2009,Baalrud_2011} have been used to study Hall MHD reconnection (with or without finite resisitivity) while using axisymmetric magnetic geometry $(r,z,\varphi)$.

Returning to the 2D self-similar XMHD model \eqref{eq:XMHD_coef}, the remaining XMHD equations \eqref{eq:Vz_dot} and \eqref{eq:psi_dot}-\eqref{eq:Bz_dot} yield the five coupled ordinary differential equations \cite{Litvinenko_2015b}
\begin{eqnarray}
\dot{\imath} - 2\gamma\,\imath & = & 2b\,(\imath - \delta\;u), \label{eq:i_math} \\
\dot{\jmath} + 2\gamma\,\jmath & = & -\;2b\,(\jmath + \delta\;v), \label{eq:j_math} \\
\dot{u} - 2\gamma\,u & = & -\;2\,b\,\imath, \label{eq:u} \\
\dot{v} + 2\gamma\,v & = & -\;2\,b\,\jmath, \label{eq:v} \\
\dot{b} & = & -\,4\;(\imath\,v \;+\; \jmath\,u), \label{eq:b}
\end{eqnarray}
where a dot represents a time derivative and $\gamma$ appears as an unconstrained time-dependent coefficient. We note that Eqs.~\eqref{eq:i_math}-\eqref{eq:v} can be combined to yield $(\dot{\imath} + \dot{u}) = 2\,\gamma\,(\imath + u) - 2\,\delta\,b\,u$ and
$(\dot{\jmath} - \dot{v}) = -\,2\,\gamma\,(\jmath - v) - 2\,\delta\,b\,v$, which can be obtained from Eq.~\eqref{eq:V_psi}, while Eq.~\eqref{eq:b} can be obtained from Eq.~\eqref{B_phi}.

While the Hall parameter ${\sf d}_{i}$ is explicitly displayed in previous works \cite{Litvinenko_2007,Litvinenko_2015a,Litvinenko_2015b,Janda_2018,Brizard_2019,Janda_2019} in order to highlight the role of the Hall term in the existence of finite-time singularities, it has been scaled out here (i.e., the characteristic length scale is now ${\sf d}_{i}L$), with only the electron-proton mass ratio parameter ${\sf d}_{e}^{2}/{\sf d}_{i}^{2} = \delta$ remaining. Hence, the corresponding dissipationless HMHD equations are obtained from Eqs.~\eqref{eq:i_math}-\eqref{eq:b} by setting $\delta = 0$. Finally, we note that, if dissipative effects (e.g., viscosity and resistivity) are retained in the 2D four-field equations \eqref{eq:Vz_dot}-\eqref{eq:Bz_dot}, it can be shown \cite{Litvinenko_2015a,Litvinenko_2015b,Janda_2018} that the self-similar XMHD model \eqref{eq:XMHD_coef} can be modified to yield the same coupled ordinary differential equations \eqref{eq:i_math}-\eqref{eq:b}.

\subsection{Normalized XMHD equations}

In Eqs.~\eqref{eq:i_math}-\eqref{eq:b}, we note that  the free coefficient $\gamma$ can actually be absorbed as an integrating factor into the definitions
\begin{equation}
\left( \begin{array}{c}
\imath(t) \\
\jmath(t)  \\
u(t)  \\
v(t) 
\end{array}\right) \;\equiv\; \left( \begin{array}{c}
I(t) \;e^{2\Gamma(t)} \\
J(t) \;e^{-2\Gamma(t)} \\
U(t) \;e^{2\Gamma(t)} \\
V(t) \;e^{-2\Gamma(t)}
\end{array}\right),
\label{eq:gamma_IJUV}
\end{equation}
where $\Gamma(t) \equiv \int_{0}^{t}\gamma(t')\;dt'$ and the initial conditions $(\imath_{0},\jmath_{0},u_{0},v_{0}) = (I_{0},J_{0},U_{0},V_{0})$ are independent of $\gamma(t)$. In addition, if the coefficient $\gamma$ is well behaved (i.e., it doesn't have singularities of its own), the exponential growth (or decay) associated with $\Gamma(t)$ reaches infinity (or zero) only as $t \rightarrow \infty$. Since we are interested in exploring the finite-time singularities of the XMHD equations \eqref{eq:i_math}-\eqref{eq:b}, we will remove $\gamma$ by inserting the transformation \eqref{eq:gamma_IJUV} into Eqs.~\eqref{eq:i_math}-\eqref{eq:b}, which yields the modified XMHD equations
\begin{eqnarray}
\dot{I} & = & 2\,b\,(I - \delta\;U), \label{eq:I_dot} \\
\dot{J} & = & -\;2\,b\,(J + \delta\;V), \label{eq:J_dot} \\
\dot{U}  & = & -\;2\,b\,I, \label{eq:U_dot} \\
\dot{V} & = & -\;2\,b\,J, \label{eq:V_dot} \\
\dot{b} & = & -\,4\;(I\;V \;+\; J\;U). \label{eq:b_dot}
\end{eqnarray}
We note that the transformation \eqref{eq:gamma_IJUV} has left Eq.~\eqref{eq:b} for $b(t)$ invariant.

These equations have three quadratic conservation laws $d{\sf C}_{k}/dt = 0$ ($k = 1,2,3$):
\begin{eqnarray}
{\sf C}_{1} &=& I\;J \;-\; \frac{\delta}{4}\;b^{2}, \label{eq:C_1} \\
{\sf C}_{2} &=& U\;V \;-\; \frac{1}{4}\;b^{2}, \label{eq:C_2} \\
{\sf C}_{3} &=& (I + U)\;(J - V) \;-\; \frac{\delta}{4}\;b^{2} \;\equiv\; {\sf C}_{1} \;-\; {\sf C}_{2} \;+\; J\,U \;-\; I\,V \;-\; \frac{1}{4}\,b^{2}, \label{eq:C_3} 
\end{eqnarray} 
whose values are determined from the initial conditions $(I_{0},J_{0},U_{0},V_{0},b_{0})$. We note that the HMHD equations are a subset of the XMHD equations \eqref{eq:I_dot}-\eqref{eq:b_dot} by discarding electron inertia ($\delta = 0$).

\begin{figure}
\epsfysize=2in
\epsfbox{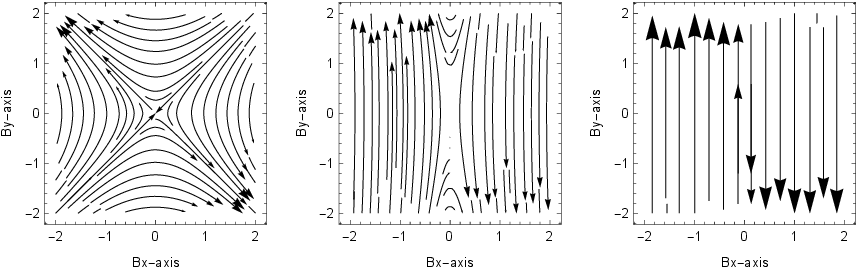}
 \caption{Plots of a generic XMHD X-point collapse in the perpendicular plane $B_{y}(x,y,t) = -2\,x\,I(t)$ versus $B_{x}(x,y,t) = -2\,y\,J(t)$ at three different times: $t = 0$ (left), where $I_{0} = J_{0}$; $t = T_{\infty}/2$ (center), where $I(t) \gg J(t)$; and $t = T_{\infty}$ (right), where $\lim_{t \rightarrow T_{\infty}}I(t) = \infty$ and $\lim_{t \rightarrow T_{\infty}}J(t) = {\cal O}(\delta)$. See Fig.~\ref{fig:ij_collapse} for additional details.}
 \label{fig:collapse}
\end{figure}

\begin{figure}
\epsfysize=1.5in
\epsfbox{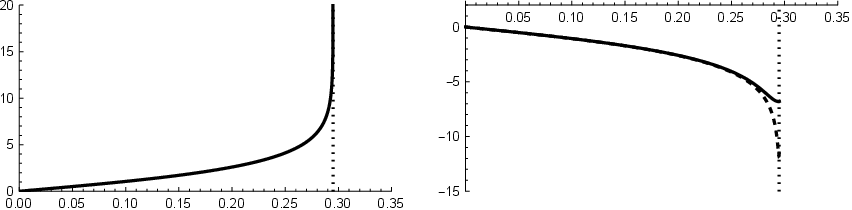}
 \caption{Plots of $\ln I(t)$ (left) and $\ln J(t)$ (right) in the range $0 \leq t < 0.35$ obtained from the numerical solutions of Eqs.~\eqref{eq:I_dot}-\eqref{eq:b_dot} for the initial conditions $(I_{0},J_{0},U_{0},V_{0},b_{0}) = (1,1,1,-1,5)$. (Left) The plot of $\ln I(t)$ clearly shows that $I(t)$ explodes exponentially at a finite time as $t \rightarrow T_{\infty}$ (indicated by the vertical dotted line). (Right) The plot of $\ln J(t)$ shows that $J(t)$ remains finite as $t \rightarrow T_{\infty}$ (solid curve). The theory presented in this paper not only predicts the correct value $T_{\infty}$ for the finite-time singularity, but also predicts the finite electron-inertia correction $\lim_{t \rightarrow T_{\infty}}J(t) = \ov{J}(\delta)$ leading to the dashed curve for $\ln[J(t) - \ov{J}(\delta)]$, which diverges as $t \rightarrow T_{\infty}$.}
 \label{fig:ij_collapse}
\end{figure}

The perpendicular magnetic-field geometry$(B_{x} = \partial\psi/\partial y,\,B_{y} = -\,\partial\psi/\partial x)$ has magnetic lines that obey the constraint $I(t)\,x^{2} - J(t)\,y^{2} = {\rm constant}$ \cite{PTR_1995}, which exhibits an X-point at the origin $(x,y)=(0,0)$ at $t = 0$ if $I_{0}\,J_{0} > 0$, or an O-point  if $I_{0}\,J_{0} < 0$. Figure \ref{fig:collapse} shows the collapse of a magnetic X-point, using the solutions for $I(t)$ and $J(t)$, based on the initial conditions $(I_{0},J_{0},U_{0},V_{0},b_{0}) = (1,1,1,-1,5)$. Here, we see the magnetic-field lines in the plane $(B_{x},B_{y})$ at three different times. At $t = 0$ (left), there is a magnetic X-point at $(x,y) = (0,0)$, while as time progresses (center) toward the finite time $t = T_{\infty}$ (right), the component $B_{x}(t)$ progressively vanishes relative to the component $B_{y}(t)$. Figure \ref{fig:ij_collapse} (left), on the other hand, shows that the numerical solution $\ln I(t)$ explodes exponentially as $t \rightarrow T_{\infty}$, while Fig.~\ref{fig:ij_collapse} (right) shows that $\ln J(t)$ reaches a finite value (solid curve), which is shown to be a finite electron-inertia correction (dashed curve). In the present work, we will derive explicit expressions for the finite singularity time $T_{\infty}$ as well as finite electron-inertia corrections for $J(t)$.

We now follow previous works \cite{Litvinenko_2007,Litvinenko_2015a,Litvinenko_2015b,Janda_2018,Brizard_2019,Janda_2019} by taking the time derivative of Eq.~\eqref{eq:b_dot}, and substituting Eqs.~\eqref{eq:I_dot}-\eqref{eq:V_dot}, to obtain a second-order ordinary differential equation for $b(t)$:
\begin{equation} 
\ddot{b}(t) \;=\; 2\;(1 + 4\,\delta)\;b^{3}(t) \;+\; 8\left[{\sf C}_{1} + (1 + 2\delta)\frac{}{}{\sf C}_{2} + {\sf C}_{3}\right] b(t) \;\equiv\; 2\;(1 + 4\,\delta)\;b^{3}(t) \;-\; 2\,{\sf C}_{0}\;b(t),
\label{eq:b_dotdot}
\end{equation}
where the constants $({\sf C}_{1},{\sf C}_{2},{\sf C}_{3})$ are defined in Eqs.~\eqref{eq:C_1}-\eqref{eq:C_3}, and the constant ${\sf C}_{0}$ is defined as
\begin{equation}
{\sf C}_{0}(I_{0},J_{0},U_{0},V_{0},b_{0}; \delta) \;\equiv\; (1 + 4\delta)\;b_{0}^{2} \;+\; 4\,\left[ I_{0}\,(V_{0} - J_{0}) \;-\frac{}{} J_{0}\,(U_{0} + I_{0}) \;-\; 2\delta\,U_{0}\,V_{0}\right].
\label{eq:C0_def}
\end{equation} 
Depending on the initial conditions $(I_{0},J_{0},U_{0},V_{0},b_{0})$ and the electron-ion mass ratio $\delta \ll 1$, the constant ${\sf C}_{0}$ may be positive or negative.

\subsection{Purpose of the present work and organization}

The purpose of the present work is to extend previous works \cite{Litvinenko_2015a,Litvinenko_2015b,Janda_2018,Brizard_2019,Janda_2019}, where ${\sf C}_{0} > 0$ was the only case considered, to include the case ${\sf C}_{0} < 0$. We will introduce a classification of bounded and unbounded solutions for the magnetic coefficient $b(t)$ associated with orbits in a quartic potential that are parametrized by ${\sf C}_{0}$ and the real parameter $\epsilon$. Each of these analytical solutions for $b(t)$ will be expressed in terms of Jacobi elliptic functions \cite{NIST_Jacobi,Lawden}, which will yield an explicit expression for the finite singularity time $T_{\infty}({\sf C}_{0},\epsilon)$ for the unbounded solutions as functions of $({\sf C}_{0},\epsilon)$. For the bounded (periodic) solution, the period will also be explicitly expressed as a function of $({\sf C}_{0},\epsilon)$. We will also show how these magnetic solutions can be used to obtain solutions of the Hall MHD equations (with $\delta = 0$), as well finite electron-inertia corrections, for the XMHD coefficients $(I,J,U,V)$.

The remainder of the paper is organized as follows. In Sec.~\ref{sec:q_Jacobi}, the magnetic equation \eqref{eq:b_dotdot} [Eq.~\eqref{eq:Energy_q}] is solved in terms of the Jacobi elliptic functions \cite{NIST_Jacobi,Lawden} for all possible values for $({\sf C}_{0},\epsilon)$, where $\epsilon$ is a real parameter. We note that Janda \cite{Janda_2018} has previously solved the magnetic equation in terms of the Weierstrass elliptic function \cite{Lawden,NIST_Weierstrass} for the case ${\sf C}_{0} > 0$ and $\epsilon < 1$. (See Refs.~\cite{Brizard_2019,Janda_2019} for comments on Janda's work \cite{Janda_2018}.) In the present work, we use a standard energy method to obtain a simple classification of ``particle'' orbits in a quartic potential \cite{Brizard_2009,Brizard_Lagrangian}, which allows bounded periodic solutions as well as unbounded solutions that exhibit finite-time singularities. In Sec.~\ref{sec:electron}, we explore how finite electron inertia modifies the solutions of the normalized XMHD coefficients $(I,J,U,V)$ once the magnetic coefficient $b(t)$ is solved. In Sec.~\ref{sec:numerical}, the numerical solutions of the XMHD equations \eqref{eq:I_dot}-\eqref{eq:b_dot} are obtained, with the value of the electron-to-proton mass ratio $\delta = 1/1836$. Here, the orbit classification introduced in Sec.~\ref{sec:q_Jacobi} is confirmed and the Jacobi elliptic solutions match the numerical results exactly, including the finite-time singularities predicted at times $T_{\infty}({\sf C}_{0},\epsilon)$ that are expressed in terms of complete elliptic integrals. We summarize our work in Sec.~\ref{sec:summary} and, in App.~\ref{sec:Weierstrass}, we present a systematic representation of our Jacobi elliptic solutions in terms of the Weierstrass elliptic function, which includes a complete (and extended) representation of Janda's solution \cite{Janda_2018} for the bounded and unbounded orbits expressed in terms of the Weierstrass elliptic function.

\section{\label{sec:q_Jacobi}Jacobi Elliptic Solution for the XMHD Magnetic Equation}

In the present Section, we show that the solutions for Eqs.~\eqref{eq:I_dot}-\eqref{eq:V_dot} can be determined from the solutions of the magnetic coefficient $b(t)$. Here, by introducing the integrating factors $e^{\pm 2\beta}$, where $\beta(t) \equiv \int_{0}^{t}
b(t')\,dt'$, with the transformations $I(t) \equiv {\cal I}(t)\,e^{2\beta(t)}$ and $J(t) \equiv {\cal J}(t)\,e^{-2\beta(t)}$, we obtain the coupled reduced XMHD equations
\begin{eqnarray}
\left( \begin{array}{c}
\dot{\cal I} \\
\dot{V} \end{array} \right) &=& \frac{d(e^{-2\beta})}{dt} \left( \begin{array}{c}
\delta\,U \\
 {\cal J}
 \end{array} \right) \;=\; -2\,b\,e^{-2\beta}\,\left( \begin{array}{c}
\delta\,U \\
 {\cal J}
 \end{array} \right), \label{eq:IV_dot} \\
\left( \begin{array}{c}
\dot{\cal J} \\
\dot{U} \end{array} \right) &=& -\,\frac{d(e^{2\beta})}{dt} \left( \begin{array}{c}
\delta\,V \\
 {\cal I}
 \end{array} \right) \;=\; -2\,b\,e^{2\beta}\,\left( \begin{array}{c}
\delta\,V \\
 {\cal I}
 \end{array} \right), \label{eq:JU_dot}
\end{eqnarray}
which can be expressed in vector form as
\begin{equation}
\dot{\vb{\chi}}_{\pm} \;=\; -2\,b\,e^{\pm 2\beta} \left( \begin{array}{cc}
0 & \delta \\
1 & 0 
\end{array}\right)\cdot\vb{\chi}_{\mp},
\end{equation}
where the reduced coefficient vectors $\vb{\chi}_{+} \equiv ({\cal J},U)^{\top}$ and $\vb{\chi}_{-} \equiv ({\cal I},V)^{\top}$ are the two linearly-independent solutions of their respective second ordinary differential equations
\begin{equation}
\ddot{\vb{\chi}}_{\pm} \;=\; \frac{e^{\mp 2\beta}}{b}\frac{d}{dt}\left(b\,e^{\pm 2\beta}\right) \dot{\vb{\chi}}_{\pm} \;+\; 4\,\delta\,b^{2}\;\vb{\chi}_{\pm} \;=\; \left( \frac{\dot{b}}{b} \;\pm\; 2\,b \right) \dot{\vb{\chi}}_{\pm} \;+\; 4\,\delta\,b^{2}\;\vb{\chi}_{\pm},
\label{eq:chi_pm}
\end{equation}
subject to the initial conditions $(I_{0},J_{0},U_{0},V_{0},b_{0})$. Hence, the solution $b(t)$ of the magnetic equation \eqref{eq:b_dotdot} can be used to generate solutions for the reduced coefficients $\vb{\chi}_{\pm}$.

For a given solution for $b(t)$, which is still determined from the initial conditions $(I_{0},J_{0},U_{0},V_{0},b_{0})$, the reduced XMHD equations \eqref{eq:IV_dot}-\eqref{eq:JU_dot} can be shown to satisfy the following conservation laws $d{\sf C}_{k}/dt = 0$, where
\begin{eqnarray}
{\sf C}_{1} &=& {\cal I}\;{\cal J} \;-\; \frac{\delta}{4}\;b^{2}, \label{eq:C_1beta} \\
{\sf C}_{2} &=& U\;V \;-\; \frac{1}{4}\;b^{2}, \label{eq:C_2beta} \\
{\sf C}_{3} &=& \left({\cal I}\,e^{2\beta} \;+\; U\right)\;\left({\cal J}\,e^{-2\beta} \;-\; V\right) \;-\; \frac{\delta}{4}\;b^{2} \;\equiv\; \left({\sf C}_{1} - {\sf C}_{2}\right) \;+\; U\,{\cal J}\;e^{-2\beta} \;-\; V\,{\cal I}\;e^{2\beta} \;-\; \frac{1}{4}\;b^{2}, \label{eq:C_3beta} 
\end{eqnarray} 
and the initial conditions for $({\cal I}_{0},{\cal J}_{0}) = (I_{0},J_{0})$ are independent of $b(t)$. We now introduce the time-dependent Hamiltonian
\begin{equation}
{\sf H}_{b}(U,V,{\cal I},{\cal J};t)  \;\equiv\; 2\,\left({\sf C}_{1} - {\sf C}_{2} - {\sf C}_{3}\right) \;=\; \frac{1}{2}\,b^{2}(t) \;+\; 2\,\left( V\,{\cal I}\;e^{2\beta(t)} \;-\; U\,{\cal J}\;e^{-2\beta(t)}\right),
\label{eq:Ham}
\end{equation}
which satisfies $\dot{\sf H}_{b} \equiv 0$, with $\dot{b} \equiv -\,4\left(V\,{\cal I}\;e^{2\beta} + U\,{\cal J}\;e^{-2\beta}\right)$, and the time-dependent canonical 4D Poisson bracket
\begin{equation}
\left\{ F,\; G\right\}_{b} \;\equiv\; b(t) \left(\pd{F}{V}\,\pd{G}{U} \;-\; \pd{F}{U}\,\pd{G}{V}\right) \;+\; \delta\,b(t) \left(\pd{F}{\cal I}\,\pd{G}{\cal J} \;-\; \pd{F}{\cal J}\,\pd{G}{\cal I}\right),
\label{eq:Poisson}
\end{equation}
Equations \eqref{eq:IV_dot}-\eqref{eq:JU_dot}  can thus be expressed in Hamiltonian form as
\begin{eqnarray}
\left(\dot{\cal I},\; \dot{\cal J}\right) &=& \delta\,b(t) \left(\pd{{\sf H}_{b} }{\cal J},\; -\,\pd{{\sf H}_{b} }{\cal I}\right) \;\equiv\; \left(\left\{{\cal I},\;{\sf H}_{b} \right\}_{b} ,\frac{}{}  \left\{{\cal J},\;{\sf H}_{b} \right\}_{b} \right), \\
\left(\dot{U},\; \dot{V}\right) &=& b(t) \left(-\,\pd{{\sf H}_{b} }{V},\; \pd{{\sf H}_{b} }{U}\right) \;\equiv\; \left(\left\{U,\;{\sf H}_{b} \right\}_{b} ,\frac{}{}  \left\{V,\;{\sf H}_{b} \right\}_{b} \right),
\end{eqnarray}
where it is immediately clear that $({\cal I},{\cal J})$ are constants of the motion for Hall MHD ($\delta = 0$). We note that an attempt at deriving a Hamiltonian structure for the 5D XMHD model $(\imath,\jmath,u,v,b)$, with $\gamma$ appearing as a free parameter, was presented in a recent paper by Abdelhamid and Lingham \cite{Abdelhamid_Lingham_2024}, but it was later revealed \cite{Abdelhamid_Lingham_2025} that their proof of the Jacobi identity in 5D space was invalid.

\subsection{Energy method}

\begin{figure}
\epsfysize=2in
\epsfbox{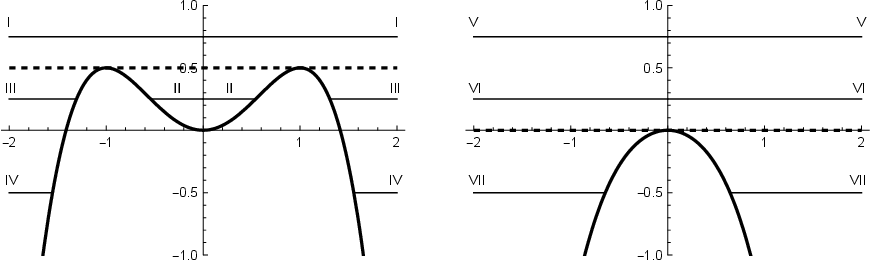}
 \caption{Potential energy $U(q)$ versus $q$ for ${\sf C}_{0} = 1$ (left) and ${\sf C}_{0} = -1$ (right). Seven different orbits (I-VII) are shown at energies $E = 3/4$ (I and V), $E = 1/4$ (II, III, and VI), and $E = -1/2$ (IV and VII). Orbit II is a bounded periodic orbit with two turning points. Orbits III, IV, and VII are unbounded orbits with a single turning point, while orbits I, V, and VI are unbounded orbits without turning points. The separatrices are shown as dashed lines: at $E = 1/2$ for the case ${\sf C}_{0} = 1$; and at $E = 0$ for the case ${\sf C}_{0} = -1$.}
 \label{fig:potential}
    \end{figure}

We now proceed with the solution $b(t)$ of the magnetic equation \eqref{eq:b_dotdot}, which can be transformed into $\ddot{q} = 2\,q^{3} - 2{\sf C}_{0}\,q$, with the definition $q(t) \equiv \sqrt{1 + 4\,\delta}\;b(t)$, which takes into account a finite electron-inertia correction. If we multiply this equation by $\dot{q}$ and integrate it over time, we obtain the energy-like conservation law \cite{Brizard_2009,Brizard_Lagrangian}
\begin{equation} 
\frac{1}{2}\,\dot{q}^{2} \;+\; U(q) \;=\; E \;=\; \frac{1}{2}\,\dot{q}_{0}^{2} \;+\; U(q_{0}) \;\equiv\; \frac{1}{2}\,{\sf C}_{0}^{2}\,\epsilon,
\label{eq:Energy_q}
\end{equation}
where the energy parameter $\epsilon$ is any real number and the ``potential'' energy is defined by the even quartic polynomial
\begin{equation} 
U(q) \;=\; {\sf C}_{0}\,q^{2} \;-\; \frac{1}{2}\,q^{4}.
\label{eq:Uq}
\end{equation}
When ${\sf C}_{0} > 0$ (solid curve in left figure in Fig.~\ref{fig:potential}), the potential energy \eqref{eq:Uq} has two maxima ${\sf C}_{0}^{2}/2$ at $q = \pm\,\sqrt{{\sf C}_{0}}$ and a local minimum $0$ at $q = 0$. When ${\sf C}_{0} < 0$, the potential energy has a single maximum $0$ at $q = 0$  (solid curve in right figure in Fig.~\ref{fig:potential}). As seen in Fig.~\ref{fig:potential} (left), the case ${\sf C}_{0} > 0$ allows for a bounded periodic solution (Orbit II: $-\,\sqrt{{\sf C}_{0}} < q < \sqrt{{\sf C}_{0}}$) for $0 < \epsilon < 1$, with $\epsilon = 1$ corresponding to a separatrix solution. We also have unbounded orbits for $0 < \epsilon < 1$, with a single turning point for ${\sf C}_{0} > 0$ (Orbit III: $q(t) >  \sqrt{{\sf C}_{0}}$ or $q(t) < -\,\sqrt{{\sf C}_{0}}$), and without turning points for ${\sf C}_{0} < 0$ (Orbit VI). For $\epsilon > 1$, the unbounded solutions have no turning points for both cases ${\sf C}_{0} > 0$ (Orbit I) and ${\sf C}_{0} < 0$ (Orbit V). For $\epsilon < 0$, the unbounded solutions have a single turning point for both cases ${\sf C}_{0} > 0$ (Orbit IV) and ${\sf C}_{0} < 0$ (Orbit VII).

\begin{figure}
\epsfysize=2.5in
\epsfbox{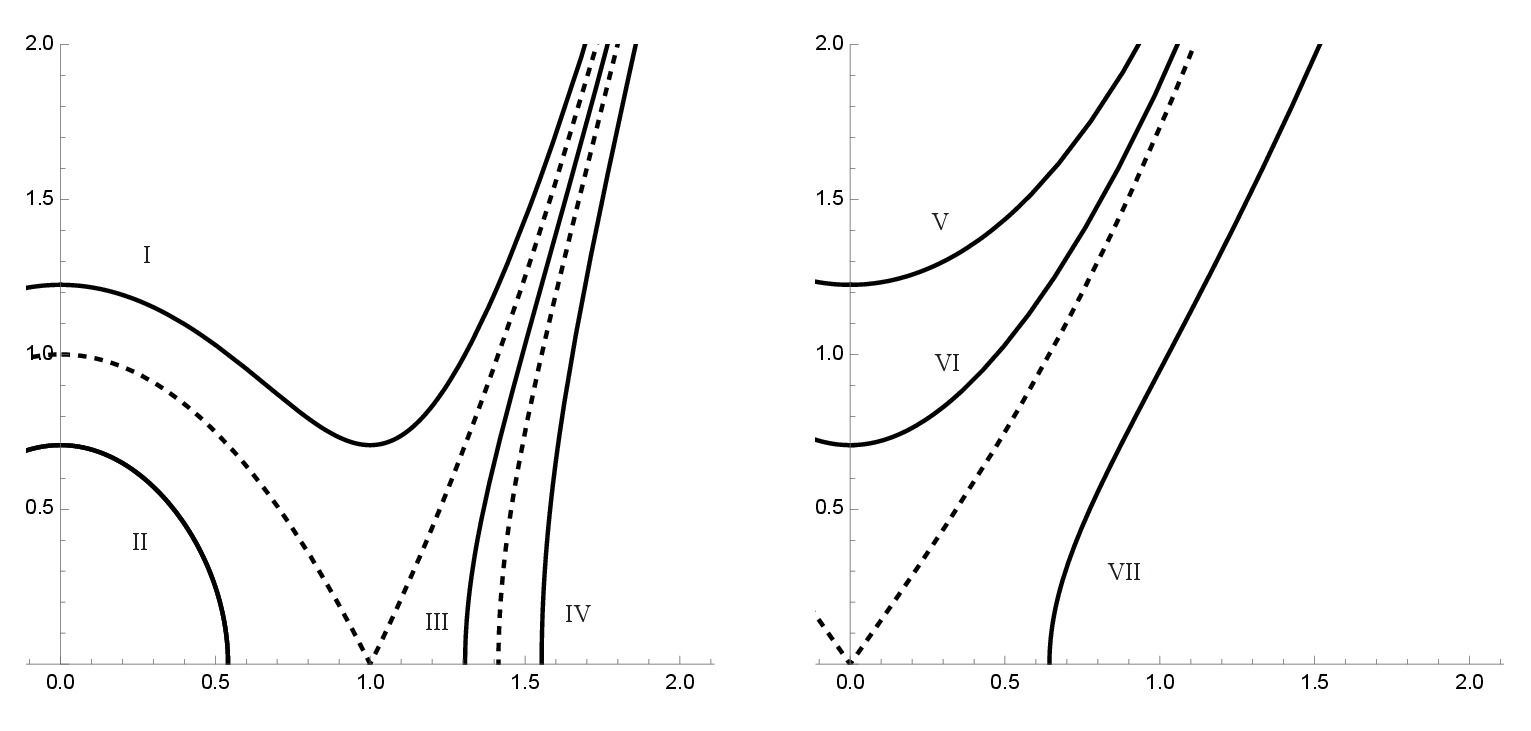}
 \caption{Phase-space portraits $(q > 0,\dot{q} > 0)$ for ${\sf C}_{0} = 1$ (left) with orbits I-IV and ${\sf C}_{0} = -1$  (right) with orbits V-VII for $\epsilon = 1.5$ (I and V), $0.5$ (II-III and VI), and $-1$ (IV and VII). The separatrices (dashed) at $\epsilon = 1$ (left) and $\epsilon = 0$ (right) are also shown.}
 \label{fig:portrait}
    \end{figure}

Equation \eqref{eq:Energy_q} may be rewritten as
\begin{equation}
\dot{q}^{2} \;=\; \left(q^{2} - {\sf C}_{0}\right)^{2} \;-\; {\sf C}_{0}^{2}\,(1 - \epsilon) \;=\; \left( q^{2} - {\sf C}_{0} - |{\sf C}_{0}|\,\sqrt{1 - \epsilon}\right)\,\left( q^{2} - {\sf C}_{0} + |{\sf C}_{0}|\,\sqrt{1 - \epsilon}\right), 
\label{eq:q_dot}
\end{equation}
whose general solutions are expressed as $q(t) = Q_{0}\;P(\Omega t)$, where $(Q_{0},\Omega)$ are functions of $({\sf C}_{0},\epsilon)$. For solutions with ${\sf C}_{0} > 0$ or ${\sf C}_{0} < 0$, respectively, the XMHD magnetic equation \eqref{eq:q_dot} becomes 
\begin{eqnarray}
(P^{\prime})^{2} &=& \frac{{\sf C}_{0}^{2}\epsilon}{Q_{0}^{2}\Omega^{2}} \left[ \frac{Q_{0}^{2}\;P^{2}}{{\sf C}_{0}\,(1 + \sqrt{1 - \epsilon})} \;-\; 1 \right]\; \left[ \frac{Q_{0}^{2}\;P^{2}}{{\sf C}_{0}\,(1 - \sqrt{1 - \epsilon})} \;-\; 1 \right],
\label{eq:P_prime_plus} \\
(P^{\prime})^{2} &=& \frac{{\sf C}_{0}^{2}\epsilon}{Q_{0}^{2}\Omega^{2}} \left[ \frac{Q_{0}^{2}\;P^{2}}{|{\sf C}_{0}|\,(1 + \sqrt{1 - \epsilon})} \;+\; 1 \right]\; \left[ 1 \;-\; \frac{Q_{0}^{2}\;P^{2}}{|{\sf C}_{0}|\,(\sqrt{1 - \epsilon} - 1)} \right].
\label{eq:P_prime_minus}
\end{eqnarray}
For orbits with finite turning points $(\dot{q} = 0)$, on the one hand, we select $q(0) = \pm\,Q_{0}$ with $P(0) = 1$ and $P^{\prime}(0) = 0$, where 
\begin{equation}
Q_{0}({\sf C}_{0},\epsilon) \;=\; \left\{ \begin{array}{lr}
\sqrt{{\sf C}_{0}(1 \pm \sqrt{1 - \epsilon})} & ({\sf C}_{0} > 0, 0 < \epsilon < 1) \\
\sqrt{{\sf C}_{0}(1 + \sqrt{1 - \epsilon})} & ({\sf C}_{0} > 0, \epsilon < 0) \\
\sqrt{|{\sf C}_{0}|(\sqrt{1 - \epsilon} - 1)} & ({\sf C}_{0} < 0, \epsilon < 0)
\end{array} \right.
\end{equation}
For orbits without turning points, on the other hand, we select $q(0) = 0$ and $\dot{q}(0) = Q_{0}\Omega$, with $P(0) = 0$ and $P^{\prime}(0) = 1$. The phase-space portraits $(q,\dot{q})$ for the cases ${\sf C}_{0} > 0$ (left) and ${\sf C}_{0} < 0$ (right) are shown in Fig.~\ref{fig:portrait}. 

The separatrix solutions for $\epsilon = 1$ $({\sf C}_{0} > 0)$ and $\epsilon = 0$ $({\sf C}_{0} < 0)$ are shown as dashed curves in Fig.~\ref{fig:portrait}. For the former separatrix, with the initial condition $q(0) = 0$ and $\dot{q} > 0$, the solution to the differential equation $\dot{q}^{2} = (q^{2} - {\sf C}_{0})^{2}$ is $q(t) \;=\; \sqrt{{\sf C}_{0}}\;{\rm tanh}(\sqrt{{\sf C}_{0}}\,t)$, which yields $\lim_{t\rightarrow\infty}q(t) = \sqrt{{\sf C}_{0}}$. For the latter separatrix, with the initial $q(0) = \sqrt{2|{\sf C}_{0}|}$ and $\dot{q} < 0$, the solution of the differential equation $\dot{q}^{2} = q^{2}\,(q^{2} + 2|{\sf C}_{0}|)$ is $q(t) = \sqrt{2|{\sf C}_{0}|}/{\rm sinh}(\sqrt{2|{\sf C}_{0}|} t + \tau_{0})$, where ${\rm sinh}\,\tau_{0} = 1$, which yields $\lim_{t\rightarrow\infty}q(t) = 0$.

\subsection{Jacobi elliptic solutions}

Next, we derive explicit solutions for the seven orbits shown in Figs.~\ref{fig:potential}-\ref{fig:portrait}. Because of the quartic potential energy \eqref{eq:Uq} is symmetric in $q^{2}$, Eqs.~\eqref{eq:P_prime_plus}-\eqref{eq:P_prime_minus} can be expressed in generic Jacobian elliptic form \cite{NIST_Jacobi,Lawden} as 
\begin{equation}
(P^{\prime})^{2} \;=\; ( \mu_{0} + \mu_{1}\,P^{2})\, ( \nu_{0} + \nu_{1}\,P^{2}), 
\label{eq:Jacobi_eq}
\end{equation}
where the parameters $(\mu_{0},\mu_{1},\nu_{0},\nu_{1})$ are functions of $({\sf C}_{0},\epsilon)$.  The Jacobi elliptic differential equations and solutions used in this work are summarized in Table \ref{tab:P_Jac}, while the orbital solutions corresponding to the orbits I-VII in Figs.~\ref{fig:potential}-\ref{fig:portrait} are summarized in Table \ref{tab:Jac_orbit}. The Jacobi elliptic functions are defined in terms of the modulus $k$, which is allowed to be complex valued under certain circumstances. In order to facilitate our discussion of the orbits considered in this work, we will consider the orbit parametrization 
\begin{equation}
\epsilon \;\equiv\; \sin^{2}\Theta \;=\; \left\{ \begin{array}{l}
\sin^{2}(\pi/2 + i\,\alpha) \;=\; \cosh^{2}\alpha \;\geq\; 1 \\
0 \;\leq\; \sin^{2}\theta \;\leq\; 1 \\
\sin^{2}(i\,\alpha) \;=\; -\;\sinh^{2}\alpha \;\leq\; 0
\end{array} \right.
\label{eq:Theta}
\end{equation}
where $\Theta = \pi/2 + i\alpha$ is complex valued when $\epsilon \geq 1$ (with $\alpha \geq 0$), $\Theta = \theta$ is real when $0 \leq \epsilon \leq 1$ (with $0 \leq \theta \leq \frac{\pi}{2}$), and $\Theta = i\alpha$ is imaginary when $\epsilon \leq 0$ (with $\alpha \geq 0$). Hence, once the energy $E = \frac{1}{2}\,\dot{q}_{0}^{2} \;+\; U(q_{0})$ of a particular orbit is determined from the initial conditions $(q_{0},\dot{q}_{0})$, with ${\sf C}_{0}$ determined from Eq.~\eqref{eq:C0_def}, we obtain a value for the real orbit parameter $\epsilon \equiv 2E/{\sf C}_{0}^{2}$, which yields an expression for $\Theta$ from Eq.~\eqref{eq:Theta}.

\begin{table}[t]
\begin{tabular}{|c|c||c|c|c|c|} \hline
$P(\tau)$                & $\;\;\;P(0)\;\;\;$           & $\;\;\;\mu_{0}\;\;\;$    & $\;\;\;\mu_{1}\;\;\;$                 & $\;\;\;\nu_{0}\;\;\;$           & $\;\;\;\nu_{1}\;\;\;$     \\ \hline
${\rm sn}(\tau, k)$     &  0                                &  1                  & $-\,1$                         & 1                            & $-\,k^{2}$    \\ \hline
${\rm sc}(\tau, k)$     &  0                               &  1                    & $1$                            & 1                           & $1 - k^{2}$    \\ \hline
${\rm nc}(\tau, k)$     &  1                               &  $-\,1$              & 1                              & $k^{2}$                  & $1 - k^{2}$    \\ \hline
${\rm cd}(\tau, k)$     &  1                               &  1                      & $-\,1$                      & 1                           & $-\,k^{2}$    \\ \hline
${\rm dc}(\tau, k)$     &  1                               &  $-\,1$              & 1                               & $-\,k^{2}$              &  1   \\ \hline
\end{tabular}
\caption{Jacobi elliptic differential equations \eqref{eq:Jacobi_eq} for orbits without turning points, with initial condition $P(0) = 0$, and orbits with turning points, with initial condition $P(0) = 1$.}
\label{tab:P_Jac}
\end{table}

\begin{table}
\begin{tabular}{|c|c|c|c|c|c|} \hline
Orbit & $\;\;\;\;\;{\sf C}_{0} \;>\; 0$                        & $Q_{0}$                                                    & $\Omega$                                                & $k^{2}$                       & $P(\Omega t)$  \\ \hline
I & $\epsilon = \cosh^{2}\alpha > 1$       & $\sqrt{{\sf C}_{0}\cosh\alpha}\,\exp(i\Phi/2)$        & $\sqrt{{\sf C}_{0}\cosh\alpha}\,\exp(-i\Phi/2)$  &  $\exp(2i\Phi)$    & ${\rm sn}(\Omega t, k)$       \\ \hline
II & $0 < \epsilon = \sin^{2}\theta < 1$        &  $\sqrt{2{\sf C}_{0}}\sin(\theta/2)$                       &  $\sqrt{2{\sf C}_{0}}\cos(\theta/2)$                     &   $\tan^{2}(\theta/2)$      &  ${\rm cd}(\Omega t, k)$      \\ \hline
III & $0 < \epsilon = \sin^{2}\theta < 1$        &  $\sqrt{2{\sf C}_{0}}\cos(\theta/2)$                       &  $\sqrt{2{\sf C}_{0}}\cos(\theta/2)$                     &   $\tan^{2}(\theta/2)$      &  ${\rm dc}(\Omega t, k)$      \\ \hline
IV & $\epsilon = -\,\sinh^{2}\alpha < 0$     &  $\sqrt{2{\sf C}_{0}}\cosh(\alpha/2)$                          &   $\sqrt{2{\sf C}_{0}\cosh\alpha}$                        &   $\sinh^{2}(\alpha/2)/\cosh\alpha$  & ${\rm nc}(\Omega t, k)$       \\ \hline \hline
 & $\;\;\;\;\;{\sf C}_{0} \;<\; 0$                         & $Q_{0}$                                                    & $\Omega$                                                & $k^{2}$                       & $P(\Omega t)$  \\ \hline 
V & $\epsilon = \cosh^{2}\alpha > 1$        &  $\sqrt{|{\sf C}_{0}|\cosh\alpha}\,\exp(i\Phi/2)$     & $\sqrt{|{\sf C}_{0}|\cosh\alpha}\,\exp(-i\Phi/2)$  & $1 -  \exp(2i\Phi)$       &   ${\rm sc}(\Omega t, k)$     \\ \hline
VI & $0 < \epsilon = \sin^{2}\theta < 1$     &  $\sqrt{2|{\sf C}_{0}|}\sin(\theta/2)$                       &  $\sqrt{2|{\sf C}_{0}|}\cos(\theta/2)$               &   $1 - \tan^{2}(\theta/2)$      &  ${\rm sc}(\Omega t, k)$      \\ \hline
VII & $\epsilon = -\,\sinh^{2}\alpha < 0$      &    $\sqrt{2|{\sf C}_{0}|}\sinh(\alpha/2)$                       &  $\sqrt{2|{\sf C}_{0}|\cosh\alpha}$                   &  $\cosh^{2}(\alpha/2)/\cosh\alpha$       &  ${\rm nc}(\Omega t, k)$      \\ \hline
\end{tabular}
\caption{Orbital solutions $q(t) = Q_{0}\,P(\Omega t)$ for the XMHD magnetic equation \eqref{eq:q_dot}, where $0 \leq \theta \leq \pi/2$ and $\alpha \geq 0$ are selected on the basis of Eq.~\eqref{eq:Theta}. For orbits I and V, we have $\Phi = \tan^{-1}(\sinh\alpha) < \pi/2$, so that $k = (1 + i\,\sinh\alpha)/\cosh\alpha \equiv \exp(i\Phi)$.}
\label{tab:Jac_orbit}
\end{table}

\subsubsection{Periodic bounded orbit}

As seen in Fig.~\ref{fig:potential} (left), there is only one bounded periodic orbit in the quartic potential \eqref{eq:Uq} with ${\sf C}_{0} > 0$. The bounded periodic orbit II has the solution $q_{II}(t) = Q_{0}\,{\rm cd}(\Omega t,k) \equiv Q_{0}\,{\rm cn}(\Omega t,k)/{\rm dn}(\Omega t, k)$, which is defined in Table \ref{tab:P_Jac}, where $q_{II}(0) = Q_{0} = \sqrt{{\sf C}_{0}(1 - \sqrt{1 - \epsilon})}$ and $0 < \epsilon = \sin^{2}\theta < 1$. The period for the regular bounded orbit II is
\begin{equation}
T_{II}({\sf C}_{0},\epsilon) \;=\; 4\,{\sf K}(k)/\Omega \;=\; 4\,\sec(\theta/2)\,{\sf K}(\tan(\theta/2))/\sqrt{2{\sf C}_{0}},
\label{eq:period_bounded}
\end{equation}
which is shown as a dashed curve in Fig.~\ref{fig:finite_time} for ${\sf C}_{0} = \frac{1}{2}$,  where \cite{NIST_EK,HMF_AS} 
\begin{equation}
{\sf K}(k) = \int_{0}^{\pi/2} d\varphi/\sqrt{1 - k^{2}\,\sin^{2}\varphi}
\label{eq:K_k}
\end{equation}
denotes the complete elliptic integral of the first kind. At the bottom of the well $(\epsilon = 0)$, the period is $T_{II}({\sf C}_{0},0) = 2\pi/\sqrt{2{\sf C}_{0}}$, while $T_{II}({\sf C}_{0},\epsilon) \rightarrow \infty$ as we approach the separatrix at $\epsilon = 1$ ($\theta = \pi/2$). 

\subsubsection{Finite-time singularity for unbounded orbits}

\begin{figure}
\epsfysize=2in
\epsfbox{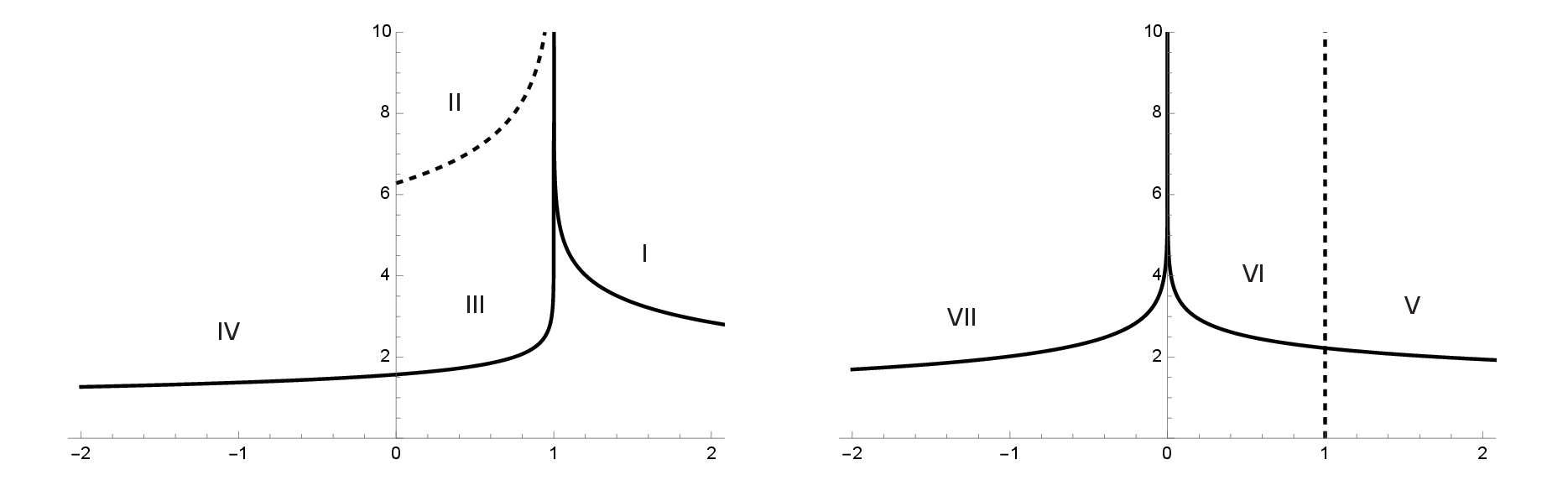}
\caption{Plots of the finite-time singularity $T_{\infty}({\sf C}_{0},\epsilon)$ (solid curves) for the unbounded singular orbits I and III-VII for ${\sf C}_{0} = 1/2$ (left) and  ${\sf C}_{0} = -\,1/2$ (right) in the range $-2 < \epsilon < 2$. The period $T(\frac{1}{2},\epsilon)$ for the regular bounded periodic orbit II in the range $0 \leq \epsilon \leq 1$ is shown as a dashed curve (left). The finite singularity times diverge at the separatrix with ${\sf C}_{0} > 0$ and $\epsilon = 1$ (left) and the separatrix with  ${\sf C}_{0} < 0$ and $\epsilon = 0$ (right).}
\label{fig:finite_time}
\end{figure}

For the unbounded orbits with a single turning point (III-IV and VII) and without a turning point (VI), the solutions $q(t) = Q_{0}\,P(\Omega t)$ reach infinity in a finite time. For orbits III-IV and VI-VII, for example, we find $q(t) \rightarrow \infty$ as $t \rightarrow T_{\infty} \equiv {\sf K}(k)/\Omega$ since ${\rm cn}({\sf K}(k),k) = 0$. The singularity finite-times $T_{\infty}({\sf C}_{0},\epsilon)$ are shown in Fig.~\ref{fig:finite_time} as solid curves for ${\sf C}_{0} > 0$ (left frame) and ${\sf C}_{0} < 0$ (right frame), and are summarized in Table \ref{tab:T}. 

\begin{table}
\begin{tabular}{|c|l|c|c|} \hline
Orbit & $\;\;\;\;\;{\sf C}_{0} \;>\; 0$                        & $P(\Omega t)$                                                                                  & $T_{\infty}({\sf C}_{0}, \epsilon)$ \\ \hline
I       & $\epsilon = \cosh^{2}\alpha > 1$          & ${\rm sn}\left(\sqrt{{\sf C}_{0}\cosh\alpha}\,e^{-i\Phi/2}\,t, e^{i\Phi}\right)$  &  ${\sf K}(\cosh(\alpha/2)/\sqrt{\cosh\alpha})/\sqrt{{\sf C}_{0}\cosh\alpha}$    \\ \hline
III      & $0 < \epsilon = \sin^{2}\theta < 1$     &  ${\rm dc}\left(\sqrt{2{\sf C}_{0}}\cos(\theta/2)\,t, \tan(\theta/2)\right)$   & $\sec(\theta/2)\,{\sf K}(\tan\theta/2)/\sqrt{2{\sf C}_{0}}$  \\ \hline
IV      &$\epsilon = -\,\sinh^{2}\alpha < 0$    & ${\rm nc}\left(\sqrt{2{\sf C}_{0}\cosh\alpha}\,t, \sinh(\alpha/2)/\sqrt{\cosh\alpha}\right)$    & ${\sf K}(\sinh(\alpha/2)/\sqrt{\cosh\alpha})/\sqrt{2{\sf C}_{0}\cosh\alpha}$  \\ \hline \hline
         & $\;\;\;\;\;{\sf C}_{0} \;<\; 0$                      & $P(\Omega t, k)$ & $T_{\infty}({\sf C}_{0}, \epsilon)$ \\ \hline
V      &   $\epsilon = \cosh^{2}\alpha > 1$        &   ${\rm sc}\left(\sqrt{|{\sf C}_{0}|\cosh\alpha}\,e^{-i\Phi/2}\,t, \sqrt{1 - e^{2i\Phi}}\right)$   & ${\sf K}(\sinh(\alpha/2)/\sqrt{\cosh\alpha})/\sqrt{|{\sf C}_{0}|\cosh\alpha}$ \\ \hline
VI     & $0 < \epsilon = \sin^{2}\theta < 1$     &  ${\rm sc}\left(\sqrt{2|{\sf C}_{0}|}\cos(\theta/2)\,t, \sqrt{1 - \tan^{2}(\theta/2)}\right)$   &  $\sec(\theta/2)\,{\sf K}(\sqrt{1 - \tan^{2}(\theta/2)})/\sqrt{2|{\sf C}_{0}|}$  \\ \hline
VII    & $\epsilon = -\,\sinh^{2}\alpha < 0$       & ${\rm nc}\left(\sqrt{2|{\sf C}_{0}|\cosh\alpha}\,t, \cosh(\alpha/2)/\sqrt{\cosh\alpha}\right)$   &  ${\sf K}(\cosh(\alpha/2)/\sqrt{\cosh\alpha})/\sqrt{2|{\sf C}_{0}|\cosh\alpha}$  \\ \hline
\end{tabular}
\caption{Finite-time singularity for unbounded orbits. The unbounded singular solutions III-IV and VI-VII become infinite as their common denominator ${\rm cn}(\Omega t, k)$ vanishes in a finite time $T_{\infty} = {\sf K}(k)/\Omega$. The finite-time singularities for the unbounded orbits I and V are calculated from the identities \eqref{eq:id_phi_1}-\eqref{eq:id_phi_2}.}
\label{tab:T}
\end{table}

We note that orbits I and V, which do not have turning points, also have real-valued finite singularity times even if $k(\Phi)$ and $\Omega(\Phi)$ are both complex valued, where $\epsilon = \cosh^{2}\alpha > 1$ and $\Phi \equiv \tan^{-1}(\sinh\alpha)$. For orbit I, which can be written as ${\rm sn}(u - iv,k)$, where $u = \sqrt{{\sf C}_{0}\cosh\alpha}\,\cos(\Phi/2)\,t$ and $v = \sqrt{{\sf C}_{0}\cosh\alpha}\,\sin(\Phi/2)\,t$, we note that ${\rm sn}(z,k)$ has a singularity at $z = 2\,{\sf K}(k) - i\,{\sf K}(k')$, which yields an expression for the finite time
\begin{equation}
|\Omega|\,e^{-i\Phi/2}\;T_{\infty} \;=\; 2\,{\sf K}(k) \;-\; i\,{\sf K}(k'),
\end{equation}
where $|\Omega|(\alpha) = \sqrt{{\sf C}_{0}\,\cosh\alpha}$ and $k^{2} = e^{2i\Phi} = 1 - k'^{2}$. Next, we introduce the identities
\begin{eqnarray}
e^{i\Phi/2}\;{\sf K}\left( e^{i\Phi}\right) &=& \frac{1}{2}\;{\sf K}\left(\cos\frac{\Phi}{2}\right) \;+\; \frac{i}{2}\;{\sf K}\left(\sin\frac{\Phi}{2}\right) \;=\;  \frac{1}{2}\;{\sf K}\left(\frac{\cosh(\alpha/2)}{\sqrt{\cosh\alpha}}\right) \;+\; \frac{i}{2}\;
{\sf K}\left(\frac{\sinh(\alpha/2)}{\sqrt{\cosh\alpha}}\right), 
\label{eq:id_phi_1} \\
e^{i\Phi/2}\;{\sf K}\left( \sqrt{1 - e^{2i\Phi}}\right) &=& {\sf K}\left(\sin\frac{\Phi}{2}\right) \;=\; {\sf K}\left(\frac{\sinh(\alpha/2)}{\sqrt{\cosh\alpha}}\right),
\label{eq:id_phi_2}
\end{eqnarray}
so that we obtain the finite singularity time for Orbit I:
\begin{equation}
|\Omega|(\alpha)\;T_{\infty}(\alpha)  \;=\; e^{i\Phi/2} \left[2\,{\sf K}\left( e^{i\Phi}\right) \;-\; i\,{\sf K}\left( \sqrt{1 - e^{2i\Phi}}\right)\right] \;=\; {\sf K}\left(\cos\frac{\Phi}{2}\right)  \;=\; {\sf K}\left(\frac{\cosh(\alpha/2)}{\sqrt{\cosh\alpha}}\right).
\label{eq:T_I}
\end{equation}
For Orbit V, on the other hand, the Jacobi elliptic solution ${\rm sc}(z,k)$ has a singularity at $z = {\sf K}(k)$, which yields $|\Omega|\;e^{-i\Phi/2}\,T_{\infty} = {\sf K}(k)$, where $|\Omega| = \sqrt{|{\sf C}_{0}|\cosh\alpha}$. Hence, using the identity \eqref{eq:id_phi_2}, we obtain the finite singularity time for Orbit V:
\begin{equation}
|\Omega|(\alpha)\;T_{\infty}(\alpha) \;=\; e^{i\Phi/2}\;{\sf K}\left( \sqrt{1 - e^{2i\Phi}}\right) \;=\; {\sf K}\left(\sin\frac{\Phi}{2}\right) \;=\; {\sf K}\left(\frac{\sinh(\alpha/2)}{\sqrt{\cosh\alpha}}\right).
\label{eq:T_V}
\end{equation}
Finally, we note that, as seen in Fig.~\ref{fig:finite_time}, the finite singularity times diverge at the separatrix with ${\sf C}_{0} > 0$ and $\epsilon = 1$ (left) and the separatrix with  ${\sf C}_{0} < 0$ and $\epsilon = 0$ (right). In addition, the solutions V-VI are smoothly connected at $\epsilon = 1$ (right): $T_{\infty} = \pi/(2\sqrt{|{\sf C}_{0}|})$.

\section{\label{sec:electron}Finite Electron Inertia XMHD Solutions}

While the electron-proton mass ratio is very small $(\delta = 1/1836 \ll 1)$, its effects cannot be entirely omitted in the solutions of the reduced XMHD equations \eqref{eq:IV_dot}-\eqref{eq:JU_dot}. In this Section, we explore how finite electron-inertia corrections can be calculated within the reduced XMHD equations, where the orbital solutions for the magnetic coefficient $b(t)$ will play a crucial role. We note that the finite singularity time $T_{\infty}({\sf C}_{0},\epsilon)$ has an implicit dependence on the electron-inertia parameter 
$\delta$ through ${\sf C}_{0}(\delta)$ and $\epsilon(\delta)$ and, in general, $dT_{\infty}/d\delta < 0$ at $\delta = 0$ (i.e., the Hall value is slightly greater than the finite-electron-inertia value).

\subsection{Hall MHD solutions}

First, we consider the numerical solution of the Hall magnetic equation $\ddot{b} = 2\,b^{3} - {\sf C}_{H}\,b$, subject to the initial conditions $(I_{0},J_{0},U_{0},V_{0},b_{0})$, with $\dot{b}_{0} = -4\,(I_{0}V_{0} + J_{0}U_{0})$, where
\begin{equation}
{\sf C}_{H} \;\equiv\; {\sf C}_{0}(\delta = 0) \;=\; b_{0}^{2} \;+\; 4\,I_{0}\left(V_{0} \;-\frac{}{} J_{0}\right) \;-\; 4\,J_{0}\left(U_{0} \;+\frac{}{} I_{0}\right), \label{eq:C0_H}
\end{equation}
so that the solution $b(t)$ is identical to $q(t)$, with the Hall parameter $\epsilon_{H} \equiv \sin^{2}\Theta_{H}$ defined as
\begin{equation}
E_{H} \;=\; \dot{b}_{0}^{2}/2 \;+\; {\sf C}_{H}\,b_{0}^{2} \;-\; b_{0}^{4}/2 \;\equiv\; \frac{1}{2}\,{\sf C}_{H}^{2}\,\epsilon_{H}. \label{eq:E_H}
\end{equation}
For orbits with a turning point, we select $b_{0} \neq 0$ and $\dot{b}_{0} = 0$, while for orbits without a turning point, we select $b_{0} = 0$ and $\dot{b}_{0} \neq 0$. 

In the absence of electron inertia $(\delta = 0)$, the reduced HMHD equations \eqref{eq:IV_dot}-\eqref{eq:JU_dot} yield the constant solutions ${\cal I}(t) = I_{0}$ and ${\cal J}(t) = J_{0}$, which can then be used to obtain the Hall solutions for Eqs.~\eqref{eq:I_dot}-\eqref{eq:V_dot}:
\begin{eqnarray}
I_{H}(t) &=& I_{0}\,e^{2\beta(t)}, \label{eq:I_0} \\
J_{H}(t) &=& J_{0}\,e^{-2\beta(t)}, \label{eq:J_0} \\
U_{H}(t) &=& U_{0} \;-\; I_{0} \left( e^{2\beta(t)} \;-\; 1 \right), \label{eq:U_0} \\
V_{H}(t) &=& V_{0} \;-\; J_{0} \left( 1 \;-\; e^{-2\beta(t)} \right), \label{eq:V_0}
\end{eqnarray}
which satisfy the Hall MHD conservation law \eqref{eq:Hall_psi}: $I_{H}(t) + U_{H}(t) = I_{0} + U_{0}$ and $J_{H}(t) - V_{H}(t) = J_{0} - V_{0}$. The solutions \eqref{eq:I_0}-\eqref{eq:V_0} of the HMHD coefficients $(I,J,U,V)$ are determined from the integral solution
\begin{equation}
\beta(t) \;=\; \int_{0}^{t}\,b(t')\,dt' \;\equiv\; \frac{Q_{0}}{\Omega}\;\int_{0}^{\Omega t} P(u)\,du,
\label{eq:beta_q}
\end{equation}
where the integral $(Q_{0}/\Omega)\int_{0}^{\Omega t}\,P(u)\,du$ for each orbit is shown in Table \ref{tab:beta}. Amazingly, all exponential factors $e^{\beta(t)}$ are expressed in terms of rational functions involving Jacobian elliptic functions, which diverge at the finite singularity times $T_{\infty}$ corresponding to each orbit in Table \ref{tab:T}.

\begin{table}
\begin{tabular}{|c|c|c|c|} \hline
Orbit        & $(Q_{0}/\Omega)$                                                                   & $P(u)$                     & $\beta(t) = (Q_{0}/\Omega)\,\int_{0}^{\Omega t}\,P(u)\,du$                      \\ \hline
I              & $e^{i\Phi} = k$                                                                           &  ${\rm sn}(u, k)$       & $\ln\left([{\rm dn}(\Omega t,k) - k\,{\rm cn}(\Omega t,k)]/\frac{}{}(1 - k)\right)$         \\ \hline
II            &  $\tan(\theta/2) = k$                                                                  &  ${\rm cd}(u, k)$      & $\ln\left({\rm nd}(\Omega t,k) \;+\frac{}{} k\,{\rm sd}(\Omega t,k)\right)$                     \\ \hline
III            &   $1$                                                                                          &  ${\rm dc}(u, k)$     & $\ln\left({\rm nc}(\Omega t,k) \;+\frac{}{} {\rm sc}(\Omega t,k)\right)$                          \\ \hline
IV           &   $\cosh(\alpha/2)/\sqrt{\cosh \alpha} = \sqrt{1 - k^{2}} = k'$      & ${\rm nc}(u, k)$       & $\ln\left({\rm dc}(\Omega t,k) \;+\frac{}{} k'\,{\rm sc}(\Omega t,k)\right)$                   \\ \hline
V            &   $e^{i\Phi} =  \sqrt{1 - k^{2}} = k'$                                              &   ${\rm sc}(u, k)$      & $\ln\left([{\rm dc}(\Omega t,k) \;+\frac{}{} k'\,{\rm nc}(\Omega t,k)]/(1 + k')\right)$   \\ \hline
VI           &   $\tan(\theta/2) = \sqrt{1 - k^{2}} = k'$                                        &  ${\rm sc}(u, k)$     &  $\ln\left([{\rm dc}(\Omega t,k) \;+\frac{}{} k'\,{\rm nc}(\Omega t,k)]/(1 + k')\right)$    \\ \hline
VII         &    $\sinh(\alpha/2)/\sqrt{\cosh \alpha} = \sqrt{1 - k^{2}} = k'$       &  ${\rm nc}(u, k)$      & $\ln\left({\rm dc}(\Omega t,k) \;+\frac{}{} k'\,{\rm sc}(\Omega t,k)\right)$                   \\ \hline
\end{tabular}
\caption{Integrals of $b(t) = Q_{0}\,P(\Omega t)$ for the HMHD magnetic equation \eqref{eq:q_dot}, where $0 \leq \theta \leq \pi/2$ and $\alpha \geq 0$.}
\label{tab:beta}
\end{table}

For example, for an unbounded orbit without a turning point (e.g., Orbit I in Table \ref{tab:beta}), we find
\begin{equation}
e^{2\beta_{I}(t)} =  \left([{\rm dn}(\Omega t,k) - k\,{\rm cn}(\Omega t,k)]/\frac{}{}(1 - k)\right)^{2\sigma},
\label{eq:beta_I}
\end{equation}
which is valid only for $0 \leq t \leq T_{\infty}$, where $T_{\infty}(\alpha)$ is defined in Eq.~\eqref{eq:T_I}. Here, both ${\rm dn}(z,k)$ and ${\rm cn}(z,k)$ have a singularity at $z = 2\,{\sf K}(k) - i\,{\sf K}(k')$, and the choice $\sigma = \pm\,1$ depends on the sign of $b(t)$ as $t \rightarrow T_{\infty}$.

For the bounded periodic orbit (Orbit II in Table \ref{tab:beta}), we find the oscillatory exponential factor
\begin{equation}
e^{2\beta_{II}(t)} = \left( {\rm nd}(\Omega t,k) \;+\frac{}{} k\,{\rm sd}(\Omega t,k) \right)^{2\sigma} \;=\; \left( \frac{1 + k\,{\rm sn}(\Omega t,k)}{{\rm dn}(\Omega t,k)}\right)^{2\sigma} \;=\;
\left( \frac{1 + k\,{\rm sn}(\Omega t,k)}{1 - k\,{\rm sn}(\Omega t,k)}\right)^{\sigma}, 
\label{eq:beta_II}
\end{equation}
which is bounded $(1 - k)/(1 + k) \leq e^{2\beta_{II}} \leq (1 + k)/(1 - k)$ for $\sigma = +1$. For an unbounded orbit with a single turning point (e.g., Orbit VII in Table \ref{tab:beta}),  we find 
\begin{equation}
e^{2\beta_{VII}(t)} = \left( {\rm dc}(\Omega t,k) \;+\frac{}{} k^{\prime}\,{\rm sc}(\Omega t,k) \right)^{2\sigma} \;=\; \left( \frac{ {\rm dn}(\Omega t,k) \;+\frac{}{} k^{\prime}\,{\rm sn}(\Omega t,k)}{{\rm cn}(\Omega t,k)}\right)^{2\sigma},
\label{eq:beta_VII}
\end{equation}
which is valid only for $0 \leq t \leq T_{\infty} = {\sf K}(k)/\Omega$, where a singularity occurs at ${\rm cn}({\sf K},k) = 0$ for $\sigma = +1$.

Lastly, if we briefly return to the original set of time-dependent coefficients $(\imath,\jmath,u,v)$ appearing in the self-similar XMHD model \eqref{eq:XMHD_coef}, the HMHD solutions \eqref{eq:I_0}-\eqref{eq:V_0} are replaced with the original HMHD solutions
\begin{eqnarray}
\imath_{H}(t) &=& I_{H}(t)\,e^{2\,\Gamma(t)} \;=\; I_{0}\,e^{2\beta(t) + 2\,\Gamma(t)}, \label{eq:I_gamma} \\
\jmath_{H}(t) &=& J_{H}(t)\,e^{-2\,\Gamma(t)} \;=\; J_{0}\,e^{-2\beta(t) - 2\,\Gamma(t)}, \label{eq:J_gamma} \\
u_{H}(t) &=& U_{H}(t)\,e^{2\,\Gamma(t)} \;=\;  (U_{0} + I_{0})\,e^{2\,\Gamma(t)} \;-\; I_{0}\,e^{2\beta(t)+ 2\,\Gamma(t)}, \label{eq:U_gamma} \\
v_{H}(t) &=& V_{H}(t)\,e^{-2\,\Gamma(t)} \;=\; (V_{0} - J_{0})\,e^{-2\,\Gamma(t)} \;+\; J_{0}\,e^{-2\beta(t) - 2\,\Gamma(t)}, \label{eq:V_gamma}
\end{eqnarray}
according to the transformation \eqref{eq:gamma_IJUV}, where $\Gamma(t) \equiv \int_{0}^{t}\gamma(t')\,dt'$. Since we have so far assumed that $\gamma(t)$ was a regular function (without finite-time singularities), we may assume that the limit $\Gamma_{\infty} \equiv \lim_{t \rightarrow T_{\infty}}\Gamma(t)$ exists. We note, here, that the evolution of $b(t)$ is unaffected by $\gamma$ and, therefore, the finite singularity time $T_{\infty}$ is also unaffected by $\gamma$. Hence, for unbounded solutions, both $\imath_{H}(t)$ and $u_{H}(t)$ diverge exponentially as $t \rightarrow T_{\infty}$, $\jmath_{H}(t) \rightarrow 0$ vanishes, while $v_{H}(t) \rightarrow (V_{0} - J_{0})\,\exp(-2\,\Gamma_{\infty})$. The most dramatic impact occurs, however, for the bounded periodic solution, where the amplitudes of oscillation may grow or decay exponentially according to the exponential factors $\exp(\pm2\,\Gamma(t))$.

\subsection{Finite electron-inertia corrections}

When $\beta(t)$ approaches infinity as $t \rightarrow T_{\infty}$ (assuming that $\sigma = +1$), the HMHD coefficients $I(t) \rightarrow \infty$ and $U(t) \rightarrow \infty$ are expected to explode exponentially, according to Eqs.~\eqref{eq:I_0} and \eqref{eq:U_0}, respectively, and, therefore, do not require finite electron inertia corrections, so that we may assume $I(t) \simeq I_{H}(t)$ and $U(t) \simeq U_{H}(t)$.  Because $\lim_{t \rightarrow T_{\infty}}V(t) = V_{0} - J_{0}$ is finite at the zeroth order in $\delta$, its finite electron inertia correction may not be important and we may again assume that $V(t) \simeq V_{H}(t)$. In the case of the Hall solution $J_{H}(t) \rightarrow 0$, however, it is important to consider corrections due to finite electron inertia, so that 
\begin{equation}
\lim_{t \rightarrow T_{\infty}}J(t) \;=\; \ov{J}(\delta) \;=\; \delta\,\ov{J}^{(1)} \;+\; \cdots 
\label{eq:J_delta}
\end{equation}
may approach a finite value at $t = T_{\infty}$. 

To derive an expression for the first-order correction $\ov{J}^{(1)}$ in Eq.~\eqref{eq:J_delta}, we begin with the expression
\begin{equation}
J(t) \;=\; J_{H}(t) \;+\; \delta\,\ov{J}^{(1)}\;\left( 1 \;-\; e^{-2\beta(t)} \right) \;+\; \cdots,
\label{eq:J1}
\end{equation}
where $J_{H}(t)$ is given by Eq.~\eqref{eq:J_0} and the initial condition $J(0) = J_{0}$ is preserved. By inserting Eqs.~\eqref{eq:I_0} and \eqref{eq:J1} into the conservation law \eqref{eq:C_1}: $I\,J - I_{0}J_{0} = \delta(b^{2} - b_{0}^{2})/4$, we obtain the first-order expression
\[
\ov{J}^{(1)}\;\left( 1 \;-\; e^{-2\beta(t)} \right) \;=\; \frac{1}{4I_{0}}\;\left(b^{2}(t) \;-\frac{}{} b_{0}^{2}\right) e^{-2\beta(t)},
\]
which is expected to be finite at $t = T_{\infty}$:
\begin{equation}
\ov{J}^{(1)} \;=\; \frac{1}{4I_{0}}\;\lim_{t \rightarrow T_{\infty}} \left(b^{2}(t) \;-\frac{}{} b_{0}^{2}\right) e^{-2\beta(t)},
\label{eq:J1_b}
\end{equation}
with $b(t)$ given in Table \ref{tab:Jac_orbit} and $e^{-2\beta(t)}$ given in Table \ref{tab:beta} in the zero electron-inertia limit $(\delta = 0)$. When an unbounded orbit has a single turning point (i.e., $b_{0} \neq 0$), for Orbit III $(0 < \epsilon < 1)$, we find 
\begin{equation}
\left(b^{2}(t) \;-\frac{}{} b_{0}^{2}\right) e^{-2\beta(t)} \;=\;  b_{0}^{2}\,\left(\frac{{\rm dn}^{2}(\Omega t,k) \;-\; {\rm cn}^{2}(\Omega t,k)}{(1 \;+\; {\rm sn}(\Omega t,k))^{2}} \right) \;\rightarrow\; \frac{b_{0}^{2}}{4}\,(1 - k^{2}) \;=\; \frac{{\sf C}_{0}}{2}\,\sqrt{1- \epsilon},
\end{equation}
while for Orbits IV and VII $(\epsilon < 0)$, we find
\begin{equation}
\left(b^{2}(t) \;-\frac{}{} b_{0}^{2}\right) e^{-2\beta(t)} \;=\;  b_{0}^{2}\,\left( \frac{{\rm sn}(\Omega t,k)}{{\rm dn}(\Omega t,k) + k^{\prime}\;{\rm sn}(\Omega t,k)}\right)^{2} \;\rightarrow\; \frac{b_{0}^{2}}{4\,(1 - k^{2})} \;=\; \frac{|{\sf C}_{0}|}{2}\,\sqrt{1- \epsilon},
\label{eq:J1_VII}
\end{equation} 
where $k = \sinh(\alpha/2)/\sqrt{\cosh\alpha}$ and  $k = \cosh(\alpha/2)/\sqrt{\cosh\alpha}$, respectively. 

For unbounded orbits without turning points (i.e., $b_{0} = 0$), on the other hand, we expect $b^{2}(t)\,e^{-2\beta(t)}$ to be finite at $t = T_{\infty}$. For Orbit I $(\epsilon > 1)$, for example, we find
\begin{eqnarray}
b^{2}(t)\,e^{-2\beta(t)} &\rightarrow& {\sf C}_{0}\,\sec\Phi\;e^{i\Phi}\,\left(1 - e^{i\Phi}\right)^{2}\;\lim_{z \rightarrow 2{\sf K}}\left( \frac{{\rm sn}(z - i\,{\sf K}',e^{i\Phi})}{{\rm dn}(z - i\,{\sf K}',e^{i\Phi}) \;-\; e^{i\Phi}\;{\rm cn}(z - i\,{\sf K}',e^{i\Phi})} \right)^{2} \nonumber \\
 &=& {\sf C}_{0}\,\sec\Phi\;e^{i\Phi}\,\left(1 - e^{i\Phi}\right)^{2}\;\lim_{z \rightarrow 2{\sf K}}\left( \frac{e^{-i\Phi}}{i\,({\rm cn}z \;-\; {\rm dn}z)} \right)^{2} \;=\; \frac{{\sf C}_{0}}{2} \left(\sqrt{\epsilon} \;-\frac{}{} 1\right),
\label{eq:I_delta}
\end{eqnarray}
where $\sec\Phi = \cosh\alpha = \sqrt{\epsilon}$, while for Orbit VI $(0 < \epsilon < 1)$, on the other hand, we find
\begin{equation}
b^{2}(t)\,e^{-2\beta(t)} \;=\; \frac{Q_{0}^{2}\,(1 + k')^{2}\;{\rm sn}^{2}(\Omega t,k)}{({\rm dn}(\Omega t,k) + k')^{2}} \;\rightarrow\; \frac{Q_{0}^{2}\,(1 + k')^{2}}{4\,k'^{2}} \;=\; \frac{1}{2}\,|{\sf C}_{0}|\;(1 +\sqrt{\epsilon}),
\end{equation}
where $k^{\prime} = \tan(\theta/2)$ and $\sin\theta = \sqrt{\epsilon}$.

Finally, we note that the present analysis would be reversed for the case $\sigma = -1$, i.e., $e^{2\beta(t)} \rightarrow 0$ as $t \rightarrow T_{\infty}$. Hence, $J(t) \rightarrow \infty$ and $V(t) \rightarrow \infty$ would explode exponentially at the finite singularity time $T_{\infty}$, $U(t) \rightarrow U_{0} + I_{0}$ would reach a finite value, and $I(t) \rightarrow \ov{I}(\delta) = \delta\,\ov{I}^{(1)} + \cdots$ would reach a finite value at first order in $\delta$, where
\begin{equation}
\ov{I}^{(1)} \;=\; \frac{-\,1}{4J_{0}}\;\lim_{t \rightarrow T_{\infty}} \left(b^{2}(t) \;-\frac{}{} b_{0}^{2}\right) e^{2\beta(t)}.
\label{eq:I1_b}
\end{equation}
This expression can, of course, also be used when $b_{0} = 0$ for orbits I and VI. We also note that, independently of the asymptotic sign of $b(t)$, i.e., whether $I(t)$ or $J(t)$ blows up exponentially at the finite-time singularity, the magnetic X-point collapse occurs.

\section{\label{sec:numerical}Numerical Solutions of the XMHD Equations}

\begin{table}
\begin{tabular}{|c|c|c|c|c|c|c|c|c|} \hline
Orbit     & $U_{0}$                     & $\;\;V_{0}\;\;$              & $I_{0}$                       & $J_{0}$                      & $b_{0}$                                     & $\;\;\dot{b}_{0}\;\;$      & $\;\;{\sf C}_{H}\;\;$  & $\epsilon_{H} = \sin^{2}\Theta_{H}$ \\ \hline
I            & $3/2$                         &  $0$                            &  $-1/4$                        &  $-1/4$                      & $0$                                            & $3/2$                         & $1$                            & $9/4 \equiv \cosh^{2}\alpha_{H}$                                      \\ \hline
II           &    $2^{-\frac{3}{4}}$     &  $-\,2^{-\frac{3}{4}}$    & $-\,2^{-\frac{7}{4}}$    & $-\,2^{-\frac{7}{4}}$    & $\sqrt{1 - 1/\sqrt{2}}$                & $0$                            &  $1$                           &  $1/2 \equiv \sin^{2}\theta_{H}$                                    \\ \hline
III          &   $0$                           &    $0$                          & $2^{-\frac{7}{4}}$        & $2^{-\frac{7}{4}}$       &  $\sqrt{1 + 1/\sqrt{2}}$             & $0$                            &  $1$                            &  $1/2 \equiv \sin^{2}\theta_{H}$                                   \\ \hline
IV          &    $0$                          &   $0$                            & $1/2$                          & $1/2$                         & $\sqrt{3}$                                & $0$                           &  $1$                             &     $-3 \equiv -\,\sinh^{2}\alpha_{H}$                                \\ \hline
V           & $\sqrt{2/3}$                 &  $0$                              &  $-\sqrt{3/8}$               &  $-\sqrt{3/8}$             & $0$                                        & $2$                           & $-1$                             & $4 \equiv \cosh^{2}\alpha_{H}$                                      \\ \hline
VI          & $1/\sqrt{12}$               &  $0$                             &  $-\sqrt{3}/4$               &  $-\sqrt{3}/4$              & $0$                                         & $1/2$                        & $-1$                           & $1/4 \equiv \sin^{2}\theta_{H}$                                      \\ \hline
VII        &   $0$                            &    $0$                           &  $1/2$                          &  $1/2$                        &  $1$                                         &  $0$                         & $-1$                             &  $-3 \equiv -\,\sinh^{2}\alpha_{H}$                                \\ \hline
\end{tabular}
\caption{Initial conditions for numerical solutions of the HMHD equations $(\delta = 0)$.}
\label{tab:initial}
\end{table}

By carefully selecting the initial conditions $(I_{0},J_{0},U_{0},V_{0},b_{0})$ from which numerical solutions of the XMHD equations \eqref{eq:I_dot}-\eqref{eq:b_dot} are obtained, we can assess how accurately the Jacobi elliptic solutions \eqref{eq:I_0}-\eqref{eq:V_0} represent the numerical solutions. The addition of the initial condition $\dot{b}_{0}$ allows us to obtain a value for the energy $E$, defined in Eq.~\eqref{eq:Energy_q}, from which we obtain the real orbit parameter $\epsilon = 2E/{\sf C}_{0}^{2} \equiv \sin^{2}\Theta$. As a general remark on the numerical solutions for $b(t)$, we note that they exactly match the Jacobi elliptic solutions $b(t) = Q_{0}\,P(\Omega t)/\sqrt{1 + 4\delta}$ for each of the orbits represented in Table \ref{tab:Jac_orbit}, where the orbit parameters $({\sf C}_{0},\epsilon)$ are calculated with full electron inertia corrections $(\delta \neq 0)$.

In this Section, we consider the cases of Orbit I (unbounded orbit without turning points), as a generic example of unbounded solutions for Eqs.~\eqref{eq:I_dot}-\eqref{eq:b_dot}, and Orbit II (bounded periodic orbit) as an example of periodic solutions for Eqs.~\eqref{eq:I_dot}-\eqref{eq:b_dot}. In order to focus on the collapse of a magnetic X-point (instead of a magnetic O-point), we select $I_{0} = J_{0}$, so that
\begin{equation}
{\sf C}_{0}(\delta) \;=\; (1 + 4\delta)\;b_{0}^{2} \;+\; 4\,\left[ I_{0}\,\left(V_{0} \;-\; U_{0} \;-\frac{}{} 2\,I_{0}\right) \;-\; 2\delta\,U_{0}\,V_{0}\right] \;\equiv\; {\sf C}_{H} \;+\; 4\,\delta \left( b_{0}^{2} \;-\frac{}{} 2\,U_{0}\,V_{0}\right),
\label{eq:C0_0}
\end{equation} 
and $\dot{b}_{0} = -4\,I_{0}\,(U_{0} + V_{0})$. Here, we have extracted the Hall contribution ${\sf C}_{H} = {\sf C}_{0}(\delta = 0)$ from the finite electron-inertia corrections. Likewise, the orbital energy can be expressed as
\begin{equation}
E(\delta) \;=\; E_{H} \;+\; 4\delta\left(\frac{1}{2}\,\dot{b}_{0}^{2} + b_{0}^{2}\,({\sf C}_{H} - 2\,U_{0}V_{0})\right) \;+\; 16\,\delta^{2} \left(\frac{1}{2}b_{0}^{2} - 2\,U_{0}V_{0}\right)\,b_{0}^{2} \;\equiv\; \frac{1}{2}\,{\sf C}_{0}^{2}(\delta)\,\epsilon(\delta),
\end{equation} 
where $E_{H} = E(\delta = 0)$ denotes the Hall orbit energy, from which we may obtain the Hall orbit parameter $\epsilon_{H} = 2\,E_{H}/{\sf C}_{H}^{2} \equiv \sin^{2}\Theta_{H}$. Table \ref{tab:initial} shows a selection of initial conditions $(I_{0},J_{0},U_{0},V_{0},b_{0}; \dot{b}_{0})$ for the Hall MHD $(\delta = 0)$ equations associated with the choices ${\sf C}_{H} = 1$ (orbits I-IV) and ${\sf C}_{H} = -1$ (orbits V-VII), as shown in Fig.~\ref{fig:potential}.

\subsection{Orbit I}

\begin{figure}
\epsfysize=2.3in
\epsfbox{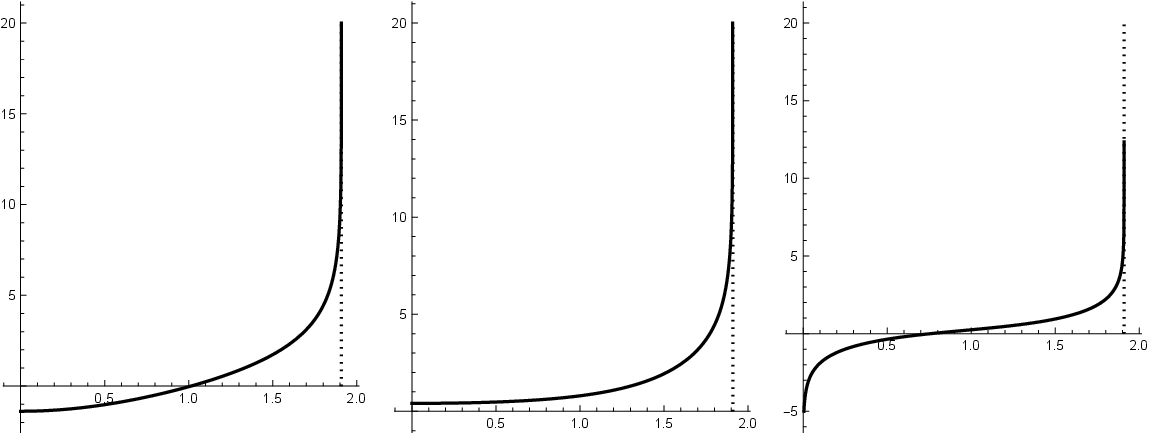}
 \caption{Numerical solutions for $\ln |I(t)|$ (left), $\ln U(t)$ (center), $\ln b(t)$ (right) versus time $0 \leq t < T_{\infty}$, where the vertical dotted line is at $t = T_{\infty}$, as defined in Table \ref{tab:T} for Orbit I.}
 \label{fig:log_IUB_I}
\end{figure}

First, we consider Orbit I (see Fig.~\ref{fig:potential}) using the initial conditions $(I_{0},J_{0},U_{0},V_{0},b_{0})$ presented in Table \ref{tab:initial}, which corresponds to the case of an unbounded orbit without a turning point. When finite electron-inertia corrections are taken into account, the HMHD orbit parameters $({\sf C}_{H}, E_{H},\epsilon_{H})$ are replaced with the XMHD parameters ${\sf C}_{I} = 1$, $E_{I} = 9(1 + 4\delta)/8$, and $\epsilon_{I} = 9(1 + 4\delta)/4 \equiv \cosh^{2}\alpha_{I}$. In Fig.~\ref{fig:log_IUB_I}, we see that the coefficients $I(t)$, $U(t)$, and $b(t)$ diverge exponentially as the time approaches the finite singularity time $t \rightarrow T_{\infty}({\sf C}_{I},\epsilon_{I}) = {\sf K}(\cosh(\alpha_{I}/2)/\sqrt{\cosh\alpha_{I}})/\sqrt{{\sf C}_{I}\,\cosh\alpha_{I}} = 1.90915$, which agrees exactly with the numerical solution for $b(t)$. We note that, as expected, $T_{\infty}({\sf C}_{I},\epsilon_{I}) < T_{\infty}({\sf C}_{H},\epsilon_{H}) = 1.91101$, i.e., the HMHD finite-time singularity occurs later than the XMHD finite-time singularity, with $T_{\infty}({\sf C}_{H},\epsilon_{H})/T_{\infty}({\sf C}_{I},\epsilon_{I}) - 1 = {\cal O}(\delta)$. 

\begin{figure}
\epsfysize=1.8in
\epsfbox{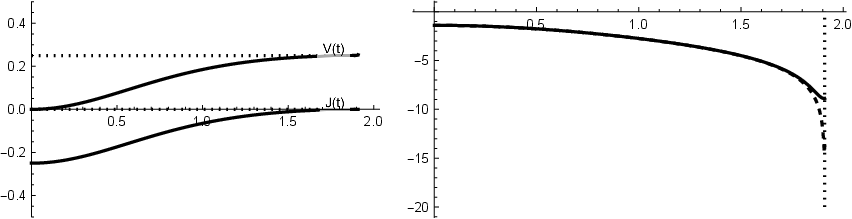}
 \caption{(Left) Numerical solutions for $V(t)$ and $J(t)$ versus time $0 \leq t < T_{\infty}$ for Orbit I, where the dotted horizontal lines show the asymptotic HMHD values. (Right) Numerical solutions for $\ln |J(t)|$ (solid curve) and $\ln|J_{H}(t)|$ (dashed curve) versus time $0 \leq t < T_{\infty}$, where $J_{H}(t) \equiv J_{0}\,e^{-2\beta(t)}$ and the vertical dotted line is at $t = T_{\infty}$, as defined in Table \ref{tab:T} for Orbit I.}
 \label{fig:VJ_I}
\end{figure}

Figure \ref{fig:VJ_I} (left) shows that $V(t)$ and $J(t)$ reach their Hall MHD limits (shown as horizontal dotted lines), $\lim_{t \rightarrow T_{\infty}}V_{H}(t) = V_{0} - I_{0} = 1/4$ and $\lim_{t \rightarrow T_{\infty}}J_{H}(t) = 0$, according to Eqs.~\eqref{eq:V_0} and \eqref{eq:J_0}, respectively. When the finite electron-inertia correction \eqref{eq:I_delta} is calculated for $J(t)$, we obtain $b^{2}(t)\,e^{-2\beta(t)} \rightarrow {\sf C}_{I}(\sqrt{\epsilon_{I}} - 1)/2$, and thus Eq.~\eqref{eq:J1_b} yields 
\begin{equation}
 \lim_{t \rightarrow T_{\infty}}J(t) \;=\; -\,\frac{1}{2}\,(\sqrt{\epsilon_{I}} - 1)\,\delta \;\equiv\; \delta\,\ov{J}_{I}^{(1)}.
 \label{eq:J_delta_I}
 \end{equation}
Figure \ref{fig:VJ_I} (right) shows that $\ln|J(t)|$ indeed reaches a finite value as $t \rightarrow T_{\infty}$ (solid curve), but by removing this finite electron-inertia correction \eqref{eq:J_delta_I}, we recover the Hall solution $J_{H}(t) = J_{0}\,e^{-2\beta(t)}$ (dashed curve).

\begin{figure}
\epsfysize=1.2in
\epsfbox{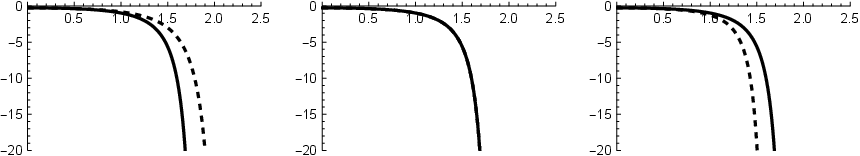}
 \caption{Comparison of the numerical solution for $I(t)$ (solid curve) and the Orbit I Jacobi elliptic solution $I(t,\alpha) =  I_{0}\,e^{2\beta_{I}(t)}$ (dashed curve) versus time $0 \leq t < T_{\infty}$, for $\alpha$: left frame ($\alpha = 0.8\,\alpha_{I}$), center frame ($\alpha = \alpha_{I}$), and right frame ($\alpha = 1.2\,\alpha_{I}$), where $\alpha_{I} \equiv {\rm \cosh}^{-1}(\sqrt{\epsilon_{I}})$.}
 \label{fig:compare_I}
\end{figure}

\begin{figure}
\epsfysize=2.2in
\epsfbox{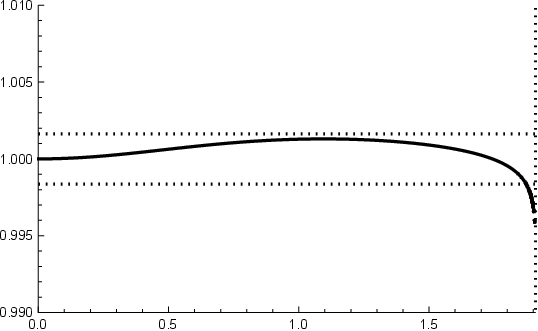}
 \caption{Plot of the ratio ${\cal I}(t)/I_{0} = I(t)/I_{H}(t)$ of the numerical solution for $I(t)$ over the Orbit I Jacobi elliptic solution $I_{H}(t,\alpha) =  I_{0}\,e^{2\beta(t)}$ versus time $0 \leq t < T_{\infty}$, for $\alpha = \alpha_{I} \equiv {\rm \cosh}^{-1}(\sqrt{\epsilon_{I}})$. The horizontal dashed lines, which are located at $1 \pm 3\,\delta$, show that the deviation of the numerical solution for $I(t)$ from the Hall solution $I_{H}(t)$ occurs mainly near the finite-time singularity at $t = T_{\infty}$.}
 \label{fig:compare_IH}
\end{figure}

We now investigate how accurately the Jacobi elliptic solution for Orbit I (see Table \ref{tab:Jac_orbit}) describe the numerical solution for $I(t)$. Figure \ref{fig:compare_I} shows the numerical solution (solid curve) of the magnetic coefficient $I(t)$ compared with the Jacobi elliptic solution (dashed curve) associated with the HMHD solution \eqref{eq:I_0}:
\begin{equation}
I_{H}(t;\alpha) \;=\;  I_{0}\,\exp(2\beta_{I}(t)) \;=\; I_{0}\left([{\rm dn}(\Omega(\alpha) t,k(\alpha)) - k(\alpha)\,{\rm cn}(\Omega(\alpha) t,k(\alpha))]/\frac{}{}(1 - k(\alpha))\right)^{2},  
\label{eq:I_Hall_I}
\end{equation}
which is evaluated for fixed ${\sf C}_{I}$, while the orbit parameter $\epsilon(\alpha) \equiv \cosh^{2}\alpha$ is allowed to vary (i.e., the energy level in Fig.~\ref{fig:potential} is allowed to vary) with the functions $\Omega(\alpha)$ and $k(\alpha)$ defined in Table \ref{tab:Jac_orbit} for Orbit I. In the center frame, the Jacobi elliptic solution \eqref{eq:I_Hall_I} matches the numerical solution $I(t)$ exactly (within 0.2 \%) when $\alpha = \alpha_{I} = \cosh^{-1}(\sqrt{\epsilon_{I}})$, which corresponds to the HMHD solution \eqref{eq:I_0} parameterized by the XMHD parameters $({\sf C}_{I},\epsilon_{I})$. We note that the other XMHD coefficients match the numerical solutions $(U,V)$ exactly  (within the same accuracy) when $\alpha \rightarrow \alpha_{I}$.

 Finally,  we show in Fig.~\ref{fig:compare_IH} that the ratio $I(t)/I_{H}(t,\alpha_{I}) \equiv {\cal I}(t)/I_{0}$ remains close to unity (the horizontal dashed lines are located at $1 \pm 3\,\delta$), until we are near the finite-time singularity at $T_{\infty}$. Indeed, using Eq.~\eqref{eq:IV_dot}, we find that the slope $\dot{\cal I}(t)$ is approximated (to first order in $\delta$) as
\[ \dot{\cal I}(t) \;\simeq\; -2\,\delta\,b(t)\,e^{-2\beta(t)}U_{H}(t) \;=\; -2\,\delta\,b(t)\,e^{-2\beta(t)}\left[U_{0} \;-\frac{}{} I_{0}\,\left( e^{2\beta(t)} \;-\; 1 \right)\right] \;\rightarrow\; 2\,\delta\,I_{0}\,b(t), \]
which deviates from zero only when $2\delta\,I_{0}\,b(t)$ becomes large (i.e., near the finite-time singularity at $T_{\infty}$).

\subsection{Orbit II}

\begin{figure}
\epsfysize=2in
\epsfbox{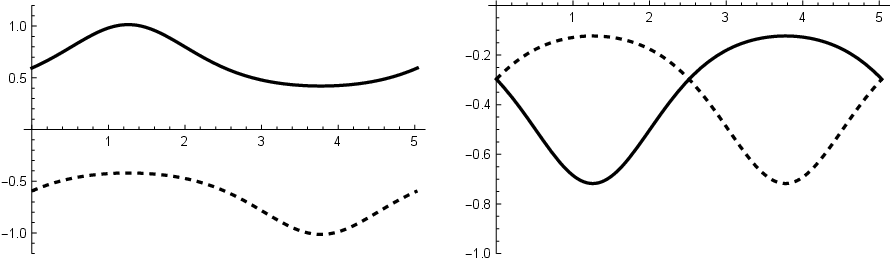}
 \caption{(Left) Plots of numerical solutions for $U(t)$ (solid) and $V(t)$ (dashed) versus time $0 \leq t \leq T_{II}$; (Right) Plots of numerical solutions for $I(t)$ (solid) and $J(t)$ (dashed) versus time $0 \leq t \leq T_{II}$.}
 \label{fig:ijuvb_II}
\end{figure}

Next, we consider the case of the bounded periodic Orbit II (see Fig.~\ref{fig:potential}), whose existence is predicted for the orbit parameters ${\sf C}_{0} > 0$ and $0 < \epsilon < 1$. Here, by choosing the initial conditions 
$(I_{0},J_{0},U_{0},V_{0},b_{0})$ shown in Table \ref{tab:initial},  the HMHD parameters $({\sf C}_{H}, E_{H}, \epsilon_{H})$ are replaced with the XMHD parameters ${\sf C}_{II} = 1 + 4\delta$, $E_{II} = \frac{1}{4}\,(1 + 4\delta)^{2}$, and $\epsilon_{II} = \frac{1}{2} \equiv \sin^{2}\theta_{II}$ (i.e., $\theta_{II} = \pi/4$). Figure \ref{fig:ijuvb_II} shows the numerical solutions for $(I,J,U,V,b)$ for one period, which is matched exactly by the Jacobi period $T_{II}({\sf C}_{II},\epsilon_{II}) = 4 \sec(\theta_{II}/2)\,{\sf K}(\tan(\theta_{II}/2))/\sqrt{2{\sf C}_{II}}= 5.032375$, defined in Eq.~\eqref{eq:period_bounded}. Once again, we note that $T_{II}({\sf C}_{II},\epsilon_{II}) < T_{II}({\sf C}_{H},\epsilon_{H}) = 5.03785$, i.e., the HMHD orbital period is slightly longer than the XMHD orbital period, with $T_{II}({\sf C}_{H},\epsilon_{H})/T_{II}({\sf C}_{II},\epsilon_{II}) - 1 = {\cal O}(\delta)$.

\begin{figure}
\epsfysize=1.2in
\epsfbox{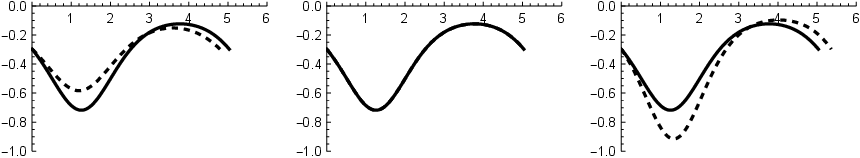}
 \caption{Comparison of the numerical solution for $I(t)$ (solid curve) and the Orbit II Jacobi elliptic solution $I_{H}(t,\theta) =  I_{0}\,e^{2\beta_{II}(t)}$ (dashed curve) versus time $0 \leq t \leq T_{II}$, for $\theta$: left frame ($\theta = 0.8\,\theta_{II}$), center frame ($\theta = \theta_{II}$), and right frame  ($\theta = 1.2\,\theta_{II}$).}
 \label{fig:compare_II}
\end{figure}

\begin{figure}
\epsfysize=2.2in
\epsfbox{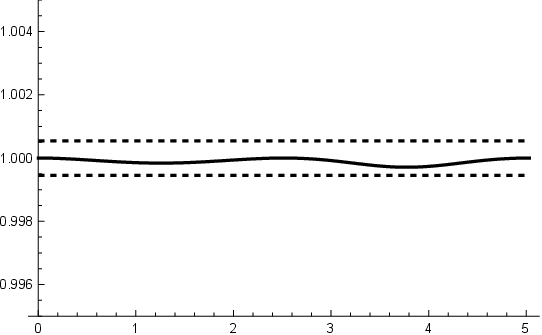}
 \caption{Plot of the ratio ${\cal I}(t)/I_{0} = I(t)/I_{H}(t)$ of the numerical solution for $I(t)$ over the Orbit II Jacobi elliptic solution $I_{H}(t,\theta) =  I_{0}\,e^{2\beta_{II}(t)}$ versus time $0 \leq t < T_{II}$, for $\theta = \theta_{II} \equiv \sin^{-1}(\sqrt{\epsilon_{II}})$. The horizontal dashed lines, which are located at $1 \pm \delta$, show that the deviation of the numerical solution for $I(t)$ from the Hall solution $I_{H}(t;\theta_{II})$ is bounded.}
 \label{fig:compare_IH_II}
\end{figure}

Figure \ref{fig:compare_II} shows the numerical solution (solid curve) of the magnetic coefficient $I(t)$ compared with the Jacobi elliptic solution (dashed curve) associated with the HMHD solution \eqref{eq:I_0}:
\begin{equation}
I_{H}(t; \theta) =  I_{0}\,e^{2\beta_{II}(t)} \;=\; I_{0}\,\left( \frac{1 + k(\theta)\,{\rm sn}(\Omega(\theta) t,k(\theta))}{1 - k(\theta)\,{\rm sn}(\Omega(\theta) t,k(\theta))}\right), 
\label{eq:I_Hall_II}
\end{equation}
which is evaluated for fixed ${\sf C}_{II}$, while the orbit parameter $\epsilon(\theta) \equiv \sin^{2}\theta$ is allowed to vary (i.e., the energy level in Fig.~\ref{fig:potential} is allowed to vary) with the functions $\Omega(\theta)$ and $k(\theta)$ defined in Table \ref{tab:Jac_orbit} for Orbit II. In the center frame, the Jacobi elliptic solution \eqref{eq:I_Hall_II} matches the numerical solution $I(t)$ exactly (within 0.03\%) when $\theta = \theta_{II} = \sin^{-1}(\sqrt{\epsilon_{II}})$. In Fig.~\ref{fig:compare_IH_II}, on the other hand, we see that the ratio $I(t)/I_{H}(t,\alpha_{I}) \equiv {\cal I}(t)/I_{0}$ remains close to unity (the horizontal dashed lines are located at $1 \pm \delta$) is very close to unity for the entire period.  We also note that the other XMHD coefficients \eqref{eq:I_0}-\eqref{eq:V_0}, with $\exp(\pm 2\beta)$ given by Eq.~\eqref{eq:beta_II}, match the numerical solutions $(J,U,V)$ exactly (within the same accuracy) in the limit when $\theta \rightarrow \theta_{II}$.

\section{\label{sec:summary}Summary}

In the present paper, we considered the analytical and numerical solutions for the five coupled ordinary differential equations \eqref{eq:I_dot}-\eqref{eq:b_dot} associated with a 2D self-similar model of the incompressible extended MHD equations. One of the major results of our work was the crucial role played by the {\it guide} magnetic field $B_{z}$, represented by the time-dependent coefficient $b(t)$, which controlled the timing of the finite-time magnetic collapse. Here, we separated the equation for the magnetic coefficient $b(t)$, whose numerical solution was exactly expressed in terms of the Jacobian elliptic functions in Sec.~\ref{sec:q_Jacobi}. These Jacobi elliptic solutions were derived from an orbit classification based on the problem of particle motion in a quartic potential with or without a local minimum, depending on the sign of ${\sf C}_{0}$. For each unbounded orbit, we were able to calculate the time $T_{\infty}$ associated with each finite-time singularity at which the magnetic X-point collapse occurs. In our work, we have shown that the effects of finite electron inertia $(\delta \neq 0)$ is to reduce the finite singularity time $T_{\infty}(\delta)$ compared to the Hall MHD result $T_{\infty}(0)$, where $T_{\infty}(0)/T_{\infty}(\delta) - 1 \sim {\cal O}(\delta)$.

In Sec.~\ref{sec:electron}, we computed the finite electron-inertia correction for either $J(t)$ or $I(t)$, depending on which coefficient reaches zero at the finite-time singularity when finite electron-inertia effects are ignored. In Sec.~\ref{sec:numerical}, the orbital classification introduced in Sec.~\ref{sec:q_Jacobi} was confirmed by numerically solving the five coupled ordinary differential equations \eqref{eq:I_dot}-\eqref{eq:b_dot}. Here, not only is the Jacobian elliptic solution for each orbit verified numerically, but the prediction of the finite-time singularity is also verified numerically.  Finally, App.~\ref{sec:Weierstrass} presents a self-consistent representation of the Jacobian elliptic solutions expressed in terms of the Weierstrass elliptic function, as well as a more complete representation of the Weierstrass solution proposed by Janda \cite{Janda_2018}.

In future work, we might consider generalizations of the work presented here, by either considering other 2D XMHD models (e.g., \cite{Ottaviani_1995,Cafaro_1998,Grasso_1999,Tassi_2010} or \cite{Craig_2003a,Craig_2003b,Craig_2005}), or considering the 2D {\it anstaz} $\phi(x,y,t) \equiv \gamma(b)\,xy$, where $\gamma(b)$ is now an arbitrary function of $b(t)$, which would be inserted into the HMHD solutions \eqref{eq:I_gamma}-\eqref{eq:V_gamma}.

\appendix

\section{\label{sec:Weierstrass}Weierstrass Elliptic Representation}

Several problems in classical mechanics are solved in terms of doubly-periodic elliptic functions \cite{Brizard_2009,Brizard_Lagrangian}: for a particle moving in a cubic potential, its bounded and unbounded orbits are naturally expressed in terms of the Weierstrass elliptic functions \cite{NIST_Weierstrass}; and for a particle moving in a quartic potential, its bounded and unbounded orbits are naturally expressed in terms of the Jacobi elliptic functions \cite{NIST_Jacobi}. Based on mathematical grounds \cite{Whittaker_Watson}, however, a general elliptic function can either be expressed in terms of Weierstrass or Jacobi elliptic functions, which implies that there is a connection between these two classes of elliptic functions. In this Appendix, we will show that, while the twelve different Jacobi elliptic functions \cite{NIST_Jacobi} can be replaced by the single Weierstrass elliptic function \cite{NIST_Weierstrass}, this explicit simplicity hides an implicit complex structure of roots and singularities. 

The purpose of the present Appendix is to present a self-contained general description of the Weierstrass elliptic function $\wp(t)$, which satisfies the Weierstrass elliptic equation  \cite{NIST_Weierstrass}
\begin{equation}
\dot{\wp}^{2} \;=\; 4\,\wp^{3} \;-\; g_{2}\,\wp \;-\; g_{3} \;\equiv\; 4\,(\wp - e_{1})\,(\wp - e_{2})\,(\wp - e_{3}),
\label{eq:wp_eq}
\end{equation}
where the parameters $(g_{2},g_{3})$ are known as the lattice invariants and $(e_{1},e_{2},e_{3})$ denote the roots of the cubic polynomial $w(s) = 4\,s^{3} - g_{2}\,s - g_{3}$, i.e., $e_{1} + e_{2} + e_{3} = 0$, while $g_{2} = -4\,(e_{1}e_{2} + e_{2}e_{3} + e_{3}e_{1})$ and $g_{3} = 4\,e_{1}e_{2}e_{3}$. The assignment of these roots and their associated half-periods $(\omega_{1},\omega_{2},\omega_{3})$, where $\wp(t + 2\,\omega_{k}) = \wp(t)$ and $\wp(\omega_{k}) = e_{k}$ (for $k =$ 1, 2, or 3), is determined by a complex dependence on the lattice invariants $(g_{2},g_{3})$. Lastly, the Weierstrass elliptic function $\wp(t)$ has singularities at 0 and $(2\,\omega_{1}, 2\,\omega_{2}, 2\,\omega_{3})$, where $(2\,\omega_{1}, 2\,\omega_{3})$ are known as lattice generators \cite{NIST_Weierstrass}, with $\omega_{2} \equiv -\,\omega_{1} - \omega_{3}$.

\subsection{Cubic roots $(e_{1},e_{2},e_{3})$ and half-periods $(\omega_{1},\omega_{2},\omega_{3})$}

The cubic polynomial $w(s) = 4\,s^{3} - g_{2}\,s - g_{3}$ appearing in Eq.~\eqref{eq:wp_eq} has local extrema at $s_{0}^{\pm} = \pm\frac{1}{2}\,\sqrt{g_{2}/3} \;\equiv\; \pm\frac{1}{2}\,\varrho$, where $w(s_{0}^{+}) = -\varrho^{3} - g_{3}$ is a minimum and $w(s_{0}^{-}) = \varrho^{3} - g_{3}$ is a maximum, which are real only if $g_{2} > 0$. The three cubic roots $(e_{1},e_{2},e_{3})$ are determined from properties of the Weierstrass elliptic function $\wp(t)$ as follows. 

The solutions of the Weierstrass elliptic equation \eqref{eq:wp_eq} have real and imaginary (or complex) half-periods $(\omega_{1},\omega_{2},\omega_{3})$ connected to the cubic roots $(e_{1},e_{2},e_{3})$ according to the relation $\wp(\omega_{k}) = e_{k}$ ($k = 1,2,3$), with the periodicity relation $\wp(t + 2\,\omega_{k}) = \wp(t)$. In addition, the Weierstrass function $\wp(t)$ has poles at the lattice points $t = (0,2\,\omega_{1},2\,\omega_{3},2\,\omega_{2})$, where $\omega_{2} \equiv -\,\omega_{1} - \omega_{3}$. We note that the half-periods $(\omega_{1},\omega_{2},\omega_{3})$ are not only functions of the functions of the invariants $(g_{2},g_{3})$, but they also depend on the discriminant $\Delta = 27\,(\varrho^{6} - g_{3}^{2})$. When $g_{3}$ changes sign (at fixed $g_{2}$ and $\Delta$), we rely on the homogeneity of the Weierstrass elliptic function \cite{Lawden,NIST_Weierstrass}
\[ \wp(z; g_{2}, g_{3}) \;\equiv\; \lambda^{-2}\wp\left(\lambda^{-1}z; \lambda^{4}\,g_{2}, \lambda^{6}\,g_{3}\right), \]
to find, using $\lambda = -i$, the $g_{3}$-inversion formula 
\begin{equation}
\wp(z;g_{2},-|g_{3}|) \;=\; -\,\wp(iz; g_{2}, |g_{3}|), 
\label{eq:g3_inversion}
\end{equation}
which leaves $(g_{2},\Delta)$ invariant but changes the sign of $g_{3}$. Hence,  the half-periods become $\omega_{k}^{-} \;=\; -i\,\omega_{k}^{+}$ ($k = 1,2,3$) and Eq.~\eqref{eq:g3_inversion} yields
\begin{equation} 
e_{k}^{-} \;=\; \wp(\omega_{k}^{-};g_{2},-|g_{3}|) \;=\; -\,\wp(i\omega_{k}^{-}; g_{2}, |g_{3}|) \;=\; -\;\wp(\omega_{k}^{+};g_{2},|g_{3}|) \;=\; -\,e_{k}^{+},
\label{eq:g3_flip}
\end{equation}
i.e., the roots change sign when $g_{3}$ changes sign. In addition, if the half-period $\omega_{k}^{+}$ is real (or imaginary) for $g_{3} > 0$, then it becomes $\omega_{k}^{-} = -i\,\omega_{k}^{+}$, which is imaginary (or real) for $g_{3} < 0$.

\begin{figure}
\epsfysize=1.8in
\epsfbox{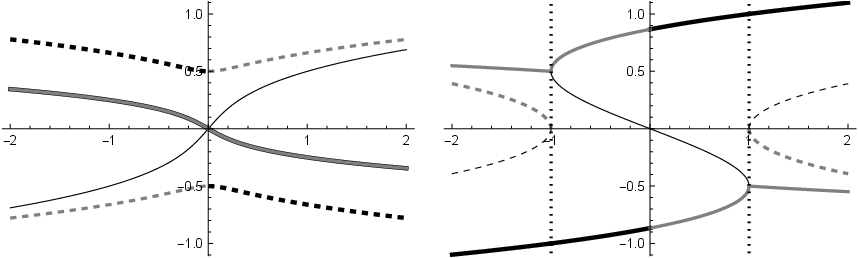}
 \caption{Plots of the real (solid) and imaginary (dashed) parts of the cubic roots $e_{1}$ (thick black curve), $e_{2}$ (light black curve), and $e_{3}$ (gray curve) in the range $-2 < g_{3} < 2$, as defined by the root assignments \eqref{eq:e123_+++}-\eqref{eq:e123_-+}. The left plot shows the case $g_{2} = -\,1 < 0$ (i.e., $\ov{\varrho} = 1/\sqrt{3}$), where the roots are defined in Eq.~\eqref{eq:e123_-+} for $g_{3} > 0$, while the right plot shows the case $g_{2} = 3 > 0$ (i.e., $\varrho = 1$), where the roots are defined in 
 Eqs.~\eqref{eq:e123_+++}-\eqref{eq:e123_++-} for $g_{3} > 0$. As $g_{3}$ changes sign, we note that each root  $e_{k}^{+}$ $(g_{3} > 0)$ is replaced with $e_{k}^{-} \equiv -\,e_{k}^{+}$ $(g_{3} < 0)$ according to the inversion formula \eqref{eq:g3_flip}. The dotted vertical lines in the right plot (at $g_{3} = \pm\,1$) show where the two roots $e_{2} = \mp\,\frac{1}{2} = e_{3}$ merge.}
 \label{fig:roots_g2}
\end{figure}

The plots of the roots  $(e_{1},e_{2},e_{3})$ are shown in Fig.~\ref{fig:roots_g2} for $g_{2} = -\,1 < 1$ (left plot) or $g_{2} = 3 > 0$ (right plot) as functions of $g_{3}$ in the range $-2 < g_{3} < 2$. Here, the real (solid) and imaginary (dashed) parts of the cubic roots are shown for $e_{1}$ as a thick black curve, for $e_{2}$ as light black curve, and for $e_{3}$ as gray curve.  The Weierstrass root assignment $(e_{1},e_{2},e_{3})$ begins with the case $g_{2} = 3\,\varrho^{2} > 0$ and $0 < g_{3} < \varrho^{3}$, i.e., $\Delta =  27\,(\varrho^{6} - g_{3}^{2}) > 0$. Here, the three roots are real and are ordered as $e_{3}^{+} < e_{2}^{+} < e_{1}^{+}$:
\begin{equation}
\left. \begin{array}{lcr}
e_{1}^{+} &=& \varrho\;\cos(\varphi/3) \\
e_{2}^{+} &=& -\,\varrho\;\cos(\pi/3 + \varphi/3) \\
e_{3}^{+} &=& -\,\varrho\;\cos(\pi/3 - \varphi/3)
\end{array} \right\},
\label{eq:e123_+++} 
\end{equation}
where $\varphi \equiv \cos^{-1}(g_{3}/\varrho^{3})$. Next, for the case $g_{2} = 3\,\varrho^{2} > 0$ and $g_{3} > \varrho^{3}$ (i.e., $\Delta < 0$), the roots are
\begin{equation}
\left. \begin{array}{lcr}
e_{1}^{+} &=& \varrho\;\cosh(\mu//3) \\
e_{2}^{+} &=& -\,\varrho\;\cos(\pi/3 + i\,\mu/3) \\
e_{3}^{+} &=& -\,\varrho\;\cos(\pi/3 - i\,\mu/3)
\end{array} \right\},
\label{eq:e123_++-} 
\end{equation}
where $\mu \equiv \cosh^{-1}(g_{3}/\varrho^{3})$, so that only $e_{1}^{+}$ is real and $e_{2}^{+} = e_{3}^{+*}$ form a complex-conjugate pair. Lastly, for the case $g_{2} = -\,3\,\ov{\varrho}^{2} < 0$, where $\Delta = -\,27\,(\ov{\varrho}^{6} + g_{3}^{2}) < 0$, the roots are
\begin{equation}
\left. \begin{array}{lcr}
e_{1}^{+} &=& -i\,\ov{\varrho}\;\cos(\pi/6 - i\,\ov{\mu}/3) \\
e_{2}^{+} &=& \ov{\varrho}\;\sinh(\ov{\mu}/3) \\
e_{3}^{+} &=& i\,\ov{\varrho}\;\cos(\pi/6 + i\,\ov{\mu}/3)
\end{array} \right\},
\label{eq:e123_-+} 
\end{equation}
where $\ov{\mu} \equiv \sinh^{-1}(g_{3}/\ov{\varrho}^{3})$ for all values of $g_{3} > 0$, so that only $e_{2}^{+}$ is real and $e_{1}^{+} = e_{3}^{+*}$ form a complex-conjugate pair for $g_{3} > 0$. In all three cases \eqref{eq:e123_+++}-\eqref{eq:e123_-+}, the roots change $e_{k}^{-} = -\,e_{k}^{+}$ according to the inversion formula \eqref{eq:g3_flip} when $g_{3}$ changes sign. 

\subsubsection{Half-periods $(\omega_{1},\omega_{2},\omega_{3})$ for $g_{2} < 0$}

\begin{figure}
\epsfysize=1.8in
\epsfbox{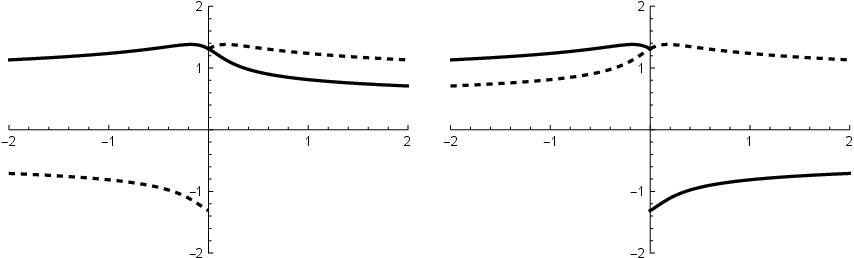}
 \caption{Plots of the half-periods $\omega_{1}$ (left) and $\omega_{3}$ (right) as functions of $g_{3}$ in the range $-2 \leq g_{3} \leq 2$ for $g_{2} = -1$. Here, the real and imaginary parts are shown as solid and dashed curves, respectively. For $g_{3} > 0$, we find $\omega_{1}^{+} = a + i\,b$ and $\omega_{3}^{+} = -a + i\,b$, where $(a,b)$ are both positive. When $g_{3} < 0$ changes sign, we have the substitutions $\omega_{1}^{-} = -\,i\,\omega_{1}^{+} = b - i\,a$ and $\omega_{3}^{-} = -i\,\omega_{3}^{+} = b + i\,a$, which follow from the relation \eqref{eq:g3_flip}.}
 \label{fig:omega_g2n}
\end{figure}

We now look at the half-periods $(\omega_{1},\omega_{3})$ for the case $g_{2} < 0$, where $e_{2}$ is real and $e_{1} = e_{3}^{*}$ form a complex-conjugate pair. The half-periods $(\omega_{1},\omega_{3})$, which are shown in Fig.~\ref{fig:omega_g2n} (for the case $g_{2} = -\,1$) and  summarized in Table \ref{tab:omega_Wn}, are defined for $g_{3} > 0$ as linear combinations of the real and imaginary integrals
\begin{eqnarray}
\wh{\omega}^{+}(g_{2},g_{3}) &\equiv& \int_{e_{2}}^{\infty}\frac{ds}{\sqrt{4s^{3} - g_{2}s - g_{3}}} \;=\; \frac{1}{\Omega}\;{\sf K}(k) \;\equiv\; 2\,a, \label{eq:omega_g2n} \\
\wh{\omega}^{\prime+}(g_{2},g_{3}) &\equiv& i\,\int_{-\infty}^{e_{2}}\frac{ds}{\sqrt{|4s^{3} - g_{2}s - g_{3}|}} \;=\; \frac{1}{\Omega} \left[2i\,{\sf K}(k') \;-\frac{}{} {\sf K}(k)\right] \;\equiv\; 2i\,b, \label{eq:omegap_g2n}
\end{eqnarray}
where ${\sf K}(k)$ and ${\sf K}(k^{\prime})$ are complete elliptic integrals, with $k^{2} = (e_{3} - e_{1})/(e_{2} - e_{1}) \equiv 1 - k^{\prime 2}$ and $\Omega \equiv \sqrt{e_{2}^{+} - e_{1}^{+}}$. Note that, while $k^{2}$ and $k^{\prime 2}$ do not change sign when $g_{3}$ changes sign, the expression $e_{2}^{-} - e_{1}^{-} = -\,(e_{2}^{+} - e_{1}^{+})$ does. Hence, for $g_{3} < 0$, we find the relations
\begin{equation}
\wh{\omega}^{-} \;=\; \frac{-i}{\Omega}\,{\sf K}(k) \;\equiv\; -i\,\wh{\omega}^{+} \;=\; -2i\,a \;\;{\rm and}\;\; \wh{\omega}^{\prime-} \;=\; \frac{-i}{\Omega}\left[2i\,{\sf K}(k') \;-\frac{}{} {\sf K}(k)\right] \;\equiv\; -i\,\wh{\omega}^{\prime+} \;=\; 2\,b,
\label{eq:omega_minus}
\end{equation}
where we wrote $\sqrt{e_{2}^{-} - e_{1}^{-}} = i\,\sqrt{e_{2}^{+} - e_{1}^{+}} \equiv i\,\Omega$. Here, we see that the half-period $\omega_{2} = -\,\omega_{1} - \omega_{3}$ is either purely imaginary $\omega_{2}^{+} = -\,\wh{\omega}^{\prime+} \equiv -2i\,b$ for $g_{3} > 0$, or purely real $\omega_{2}^{-} = -\,\wh{\omega}^{\prime-} \equiv -2\,b$ for $g_{3} < 0$.

\begin{table}
\begin{tabular}{|c|c|c|} \hline
$(g_{3})$                & $(+)$                                                                                                                                & $(-)$                                                                                                                               \\ \hline
$\omega_{1}$        & $(\wh{\omega}^{+} + \wh{\omega}^{\prime+})/2 = i\,{\sf K}'/\Omega \;\equiv\; a + i\,b$                       &  $(\wh{\omega}^{-} + \wh{\omega}^{\prime-})/2 = {\sf K}'/\Omega  \;\equiv\; b - i\,a$      \\ \hline
$\omega_{3}$        & $(-\,\wh{\omega}^{+} + \wh{\omega}^{\prime+})/2 = (i\,{\sf K}' - {\sf K})/\Omega \;\equiv\; -\,a + i\,b$     &  $(-\,\wh{\omega}^{-} + \wh{\omega}^{\prime-})/2 = ({\sf K}' + i\, {\sf K})/\Omega  \;\equiv\; b + i\,a$                     \\ \hline
\end{tabular}
\caption{Half-periods $\omega_{1} \equiv (\wh{\omega} + \wh{\omega}^{\prime})/2$ and $\omega_{3}  \equiv (-\,\wh{\omega} + \wh{\omega}^{\prime})/2$ for $g_{2} < 0$, where $\Omega \equiv \sqrt{e_{2}^{+} - e_{1}^{+}}$ and the integrals $(\wh{\omega}^{+},
\wh{\omega}^{\prime+})$ are defined by Eqs.~\eqref{eq:omega_g2n}-\eqref{eq:omegap_g2n}, respectively, while  the integrals $(\wh{\omega}^{-},\wh{\omega}^{\prime-})$ are defined in Eq.~\eqref{eq:omega_minus}.}
\label{tab:omega_Wn}
\end{table}

\subsubsection{Half-periods $(\omega_{1},\omega_{2},\omega_{3})$ for $g_{2} > 0$}

Next, we look at the half-periods $(\omega_{1},\omega_{3})$ for the case $g_{2} > 0$ and $g_{3} > 0$, which are defined, respectively, by the real and imaginary integrals
\begin{eqnarray}
\omega^{+}(g_{2},g_{3}) &\equiv& \int_{e_{1}^{+}}^{\infty}\frac{ds}{\sqrt{4s^{3} - g_{2}s - g_{3}}} \;=\; \frac{1}{\Omega}\;{\sf K}(k), \label{eq:omega_g2p} \\
\omega^{\prime+}(g_{2},g_{3}) &\equiv& i\,\int_{-\infty}^{e_{3}^{+}}\frac{ds}{\sqrt{|4s^{3} - g_{2}s - g_{3}|}} \;=\; \frac{i}{\Omega}\left[ 2\,{\sf K}(k') \;-\frac{}{} {\sf K}(k)\right], \label{eq:omegap_g2p}
\end{eqnarray}
where ${\sf K}(k)$ and ${\sf K}(k^{\prime})$ are complete elliptic integrals, with $k^{2} = (e_{2}^{+} - e_{3}^{+})/(e_{1}^{+} - e_{3}^{+}) \equiv 1 - k^{\prime 2}$ and $\Omega \equiv \sqrt{e_{1}^{+} - e_{3}^{+}}$. We note that these definitions are valid for $g_{3} > 0$ and $\Delta > 0$, where all three roots $(e_{1}^{+},e_{2}^{+},e_{3}^{+})$ are real, as well as $g_{3} > 0$ and $\Delta < 0$, where only $e_{1}^{+}$ is real and $e_{2}^{+} = e_{3}^{+*}$ form a complex-conjugate pair [see Eq.~\eqref{eq:e123_+++}], for which $(k,k')$ and $\Omega$ in Eqs.~\eqref{eq:omega_g2p}-\eqref{eq:omegap_g2p} are complex-valued. Once again, when $g_{3}$ changes sign, we find $\omega^{-} = -i\,\omega^{+}$ and $\omega^{\prime-} = -i\,\omega^{\prime+}$, where we used $\sqrt{e_{1}^{-} - e_{3}^{-}} = i\,\Omega$ while $(k,k')$ are invariant under $g_{3}$-sign inversion. 

\begin{figure}
\epsfysize=2in
\epsfbox{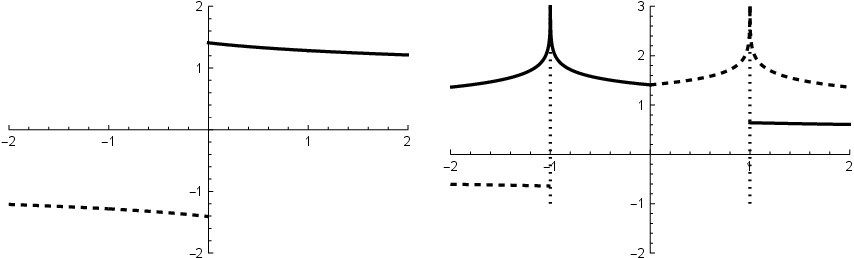}
 \caption{Plots of the half-periods $\omega_{1} \equiv \omega$ (left) and $\omega_{3} \equiv \omega^{\prime}$ (right) as functions of $g_{3}$ (in the range $-2 \leq g_{3} \leq 2$) for $g_{2} = 3$. Here, the real and imaginary parts are shown as solid and dashed curves, respectively. In addition, the vertical dotted lines at $g_{3} = \pm1$ indicate where $\Delta$ vanishes, where $\omega^{\prime}$ has a singularity since $e_{2}^{\pm} = e_{3}^{\pm}$ and $k^{2} = 0 = 1 - k^{\prime 2}$, with ${\sf K}(0) = \pi/2$ and $\lim_{k' \rightarrow 1}
 {\sf K}(k') = \infty$. When $g_{3}$ changes sign, we have the substitutions $\omega_{1}^{-} = -\,i\,\omega_{1}^{+}$ and $\omega_{3}^{-} = -i\,\omega_{3}^{+}$.}
 \label{fig:omega_13}
\end{figure}

The half-periods $\omega_{1}$ and $\omega_{3}$ are shown in Fig.~\ref{fig:omega_13} for the case $g_{2} = 3$, where $\omega_{1}^{+}$ (left) is real for $g_{3} > 0$ and imaginary $\omega_{1}^{-} = -i\,\omega_{1}^{+}$  for $g_{3} < 0$. The behavior of the half-period 
$\omega_{3}$ (right) is more intricate, as it is imaginary for $0 < g_{3} < 1$ and real for $-1 < g_{3} < 0$ (i.e., $\Delta > 0$), while it is complex valued for $|g_{3}| > 1$ (i.e., $\Delta < 0$). The half-periods $\omega_{1}$ and $\omega_{3}$ are summarized in Table \ref{tab:omega_Wp}, where the real and imaginary integrals $(\omega,\omega^{\prime})$ are defined by Eqs.~\eqref{eq:omega_g2p}-\eqref{eq:omegap_g2p}, respectively.

\begin{table}
\begin{tabular}{|c|c|c|} \hline
$(g_{3},\Delta)$     & $(+,+/-)$   & $(-,+/-)$                                                                        \\ \hline
$\omega_{1}$        & $\omega^{+} = {\sf K}/\Omega$ 
                              &  $\omega^{-} = -i\,{\sf K}/\Omega$   \\ \hline
$\omega_{3}$        & $\omega^{\prime+} = i\,(2\,{\sf K}^{\prime} - {\sf K})/\Omega $  
                              & $\omega^{\prime-}  = -i\,\omega^{\prime+} =  (2\,{\sf K}^{\prime} - {\sf K})/\Omega$                        \\ \hline
\end{tabular}
\caption{Half-periods $\omega_{1}$ and $\omega_{3}$ for $g_{2} > 0$, where the real and imaginary integrals $(\omega,\omega^{\prime})$ are defined by Eqs.~\eqref{eq:omega_g2p}-\eqref{eq:omegap_g2p}, respectively.}
\label{tab:omega_Wp}
\end{table}

\subsection{Connection between Jacobi and Weierstrass elliptic functions}

We now explore the connection between the Jacobi and Weierstrass elliptic functions. We begin with the transformation \cite{Brizard_2009}
\begin{equation}
\wp(t + t_{0}) \;=\; p_{0}\,P^{2}(\Omega t,k) \;+\; e_{0}, 
\label{eq:wp_P}
\end{equation}
where $(k,\Omega)$ and the parameter $p_{0}$ are functions of the cubic roots $(e_{1}, e_{2},e_{3})$, while $e_{0}$ is a root of the cubic polynomial $w(s) = 4\,s^{3} - g_{2}s - g_{3} = 0$, and the initial-time constant $t_{0}$ is the solution of $\wp(t_{0}) = p_{0}\,P(0)^{2} + e_{0}$. First, we substitute $\dot{\wp} = 2p_{0}\Omega\,P\,P^{\prime}$ into the Weierstrass equation $\dot{\wp}^{2} = 4\,\wp^{3} - g_{2},\wp - g_{3}$, which yields the equation
\begin{eqnarray}
\dot{\wp}^{2} \;=\; 4\,p_{0}^{2}\Omega^{2}\,P^{2}\left(P^{\prime}\right)^{2} &=& 4\left( p_{0}^{3}\,P^{6} \;+\frac{}{} 3\,p_{0}^{2}e_{0}\;P^{4} \;+\; 3\,p_{0}\,e_{0}^{2}\;P^{2} \;+\; e_{0}^{3}\right) \;-\; g_{2}\,\left(p_{0}\,P^{2} \;+\frac{}{} e_{0}\right) \;-\; g_{3}, \nonumber \\
 &=& 4\,p_{0}^{3}\;P^{6} \;+\; 12\,p_{0}^{2}e_{0}\;P^{4} \;+\; p_{0}\left( 12\,e_{0}^{2} \;-\frac{}{} g_{2}\right)\;P^{2},
 \label{eq:wp_dot2}
\end{eqnarray}
where we used the fact that $e_{0}$ is an arbitrary cubic root. Next, from Eq.~\eqref{eq:wp_dot2}, we obtain
\begin{equation}
\left(P^{\prime}\right)^{2} \;=\; \frac{p_{0}}{\Omega^{2}}\;P^{4} \;+\; \frac{3\,e_{0}}{\Omega^{2}}\;P^{2} \;+\; \left(\frac{12\,e_{0}^{2} - g_{2}}{4\,p_{0}\,\Omega^{2}}\right) \;\equiv\; \left(\mu_{0} + \mu_{1}P^{2}\right)\;\left(\nu_{0} + \nu_{1}P^{2}\right),
\end{equation}
which yields a quadratic polynomial in $P^{2}$ that can be expressed in the generic Jacobian elliptic form \eqref{eq:Jacobi_eq}, so that 
\begin{equation}
\left. \begin{array}{rcl}
p_{0} &=& \mu_{1}\nu_{1}\;\Omega^{2} \\
e_{0} &=& (\mu_{0}\,\nu_{1} + \mu_{1}\,\nu_{0})\;\Omega^{2}/3 \\
12\,e_{0}^{2} - g_{2} &=& 4\,\mu_{0}\nu_{0}\;p_{0}\,\Omega^{2}
\end{array}\right\}.
\label{eq:ab}
\end{equation}
Table \ref{tab:J_wp} shows that, by inserting the parameters $(\mu_{0},\mu_{1},\nu_{0},\nu_{1})$ associated with each of the five Jacobi elliptic functions listed in Table \ref{tab:P_Jac}, we can obtain explicit expressions \eqref{eq:ab} for $(p_{0},e_{0})$, with the generic assignment $\Omega^{2} = e_{a} - e_{c}$ and $k^{2} = (e_{b} - e_{c})/(e_{a} - e_{c})$, so that each Jacobi elliptic function $P(\Omega t, k) = \sqrt{(\wp(t+t_{0}) - e_{0})/p_{0}}$ can be expressed in terms of the Weierstrass elliptic function $\wp(t + t_{0})$ \cite{NIST_Weierstrass}.

\begin{table}
\begin{tabular}{|c|c|c|c|} \hline
Orbit        & $p_{0}$                                                   & $e_{0}$                                                    & $P(\Omega t, k) = \sqrt{(\wp(t+t_{0}) - e_{0})/p_{0}}$   \\ \hline
I               & $k^{2}\Omega^{2} = e_{b} - e_{c}$          & $-\,(1 + k^{2})\Omega^{2} = e_{c}$         & ${\rm sn}(\Omega t,k) = \sqrt{(\wp(t+\omega_{c}) - e_{c})/(e_{b} - e_{c})}$ \\ \hline
II              & $k^{2}\Omega^{2} = e_{b} - e_{c}$          & $-\,(1 + k^{2})\Omega^{2} = e_{c}$         & ${\rm cd}(\Omega t,k) = \sqrt{(\wp(t+\omega_{b}) - e_{c})/(e_{b} - e_{c})}$ \\ \hline
III             & $\Omega^{2} = e_{a} - e_{c}$                  & $-\,(1 + k^{2})\Omega^{2} = e_{c}$         & ${\rm dc}(\Omega t,k) = \sqrt{(\wp(t+\omega_{a}) - e_{c})/(e_{a} - e_{c})}$ \\ \hline
IV \& VII   & $(1 -k^{2})\Omega^{2} = e_{a} - e_{b}$   & $(2k^{2} - 1)\Omega^{2} = e_{b}$            & ${\rm nc}(\Omega t,k) = \sqrt{(\wp(t+\omega_{a}) - e_{b})/(e_{a} - e_{b})}$ \\ \hline
V \& VI    & $(1 -k^{2})\Omega^{2} = e_{a} - e_{b}$   & $(2- k^{2})\Omega^{2} = e_{a}$                & ${\rm sc}(\Omega t,k) = \sqrt{(\wp(t+\omega_{a}) - e_{a})/(e_{a} - e_{b})}$ \\ \hline
\end{tabular}
\caption{Connection between a Jacobi elliptic function $P(\Omega t,k)$ and the Weierstrass elliptic function $\wp(t + t_{0})$. The root assignment $(e_{a},e_{b},e_{c})$ depends on the signs of lattice invariants $(g_{2},g_{3})$.}
\label{tab:J_wp}
\end{table}

Lastly, we note that the generic Jacobi solution is expressed as
\begin{equation} 
q(t) \;=\; Q_{0}\,P(\Omega t,k) \;\equiv\; Q_{0} \sqrt{(\wp(t+t_{0}) - e_{0})/p_{0}} \;=\; \sqrt{\wp(t + t_{0}) \;-\frac{}{} e_{0}},
\label{eq:q_P}
\end{equation}
where $Q_{0} = \sqrt{p_{0}}$. This solution has a clear singularity at the finite time $T_{\infty} = 2\,\omega_{0} - t_{0}$, where $2\,\omega_{0} - t_{0}$ is chosen to be real. Hence, Orbit I has a singularity at $T_{\infty} = \omega_{c}$, which is chosen to be a real linear combination of $(\omega_{1},\omega_{3})$. Orbits III-VII, on the other hand, have singularities at $T_{\infty} = \omega_{a}$, which is once again chosen to be a real linear combination of $(\omega_{1},\omega_{3})$. The bounded Orbit II, on the other hand, is periodic with period $4\,\omega_{a}$, where $\omega_{a}$ is chosen to be a real linear combination of $(\omega_{1},\omega_{3})$, while the piecewise-continuous solution is
\[ q_{II}(t) \;=\; \left\{ \begin{array}{ll}
\sqrt{\wp(t + \omega_{b}) \;-\frac{}{} e_{c}} &  \hspace*{0.2in}(-\,\omega_{a} \leq t \leq \omega_{a}) \\
 & \\
-\; \sqrt{\wp(t + \omega_{b}) \;-\frac{}{} e_{c}} &  \hspace*{0.2in}(\omega_{a} \leq t \leq 3\,\omega_{a})
 \end{array} \right. \]
where $q_{II}(0) = Q_{0} = \sqrt{e_{b} - e_{c}} \equiv -\,q_{II}(2\omega_{a})$ and $q_{II}(\pm\,\omega_{a}) = 0 = q_{II}(3\omega_{a})$, where we used the even parity and periodicity properties of the Weierstrass elliptic function $\wp(t \pm 2\,\omega_{k}) = \wp(t)$, with $\omega_{b} \equiv -\,\omega_{a} - \omega_{c}$.

\subsection{Weierstrass elliptic solutions to the XMHD magnetic equation}

We now need to find expression for the cubic roots $(e_{1},e_{2},e_{3})$ and the Weierstrass invariants $(g_{2},g_{3},\Delta)$ in terms of the orbit parameters $({\sf C}_{0},\epsilon)$ used in the Jacobi elliptic solutions. For this purpose, we insert Eq.~\eqref{eq:q_P} into the magnetic differential equation $\dot{q}^{2} = q^{4} - 2\,{\sf C}_{0}q^{2} + {\sf C}_{0}^{2}\epsilon$. First, using the generic relation \eqref{eq:q_P}, we obtain
\[ \dot{q}^{2} \;=\; \frac{\dot{\wp}^{2}}{4\,(\wp - e_{0})} \;=\; (\wp - e_{0})^{2} \;-\; 2{\sf C}_{0}\,(\wp - e_{0}) \;+\; {\sf C}_{0}^{2}\epsilon, \]
which yields
\begin{eqnarray}
\dot{\wp}^{2} &=& 4\,\left(\wp^{3} \;-\frac{}{} 3\,e_{0}\,\wp^{2} \;+\; 3\,e_{0}^{2}\,\wp \;-\; e_{0}^{3}\right) \;-\; 8\,{\sf C}_{0}\,\left(\wp^{2} \;-\frac{}{} 2\,e_{0}\,\wp \;+\; e_{0}^{2}\right) \;+\; 4\,{\sf C}_{0}^{2}\,\epsilon\,(\wp - e_{0}) \nonumber \\
 &=& 4\,\wp^{3} \;-\; 4\,(3\,e_{0} + 2\,{\sf C}_{0})\,\wp^{2} \;+\; 4\,\left( 3\,e_{0}^{2} \;+\frac{}{} 4\,{\sf C}_{0}\,e_{0} \;+\; {\sf C}_{0}^{2}\epsilon\right)\,\wp \;-\; 4\,\left( e_{0}^{3} \;+\frac{}{} 2\,{\sf C}_{0}\,e_{0}^{2} \;+\; {\sf C}_{0}^{2}\epsilon\,e_{0}\right).
 \end{eqnarray}
 Since the Weierstrass equation \eqref{eq:wp_eq} does not have a $\wp^{2}$-term, we must have
 \begin{equation}
 e_{0} \;=\; -\,\frac{2}{3}\,{\sf C}_{0},
 \label{eq:e_c}
 \end{equation}
 which is always real for real values of ${\sf C}_{0}$, while the remaining terms yield
 \begin{eqnarray}
g_{2} &=& -\,4\left( 3\,e_{0}^{2} \;+\frac{}{} 4\,{\sf C}_{0}\,e_{0} \;+\; {\sf C}_{0}^{2}\epsilon\right)\ \;=\; \frac{4}{3}\,{\sf C}^{2}_{0}(4 - 3\,\epsilon), \label{eq:g_2} \\
g_{3} &=& 4\,\left( e_{0}^{3} \;+\frac{}{} 2\,{\sf C}_{0}\,e_{0}^{2} \;+\; {\sf C}_{0}^{2}\epsilon\,e_{0}\right) \;=\; \frac{8}{27}\,{\sf C}_{0}^{3}\,(8 - 9\,\epsilon). \label{eq:g_3}
\end{eqnarray}
Since $g_{2} \equiv 2\,(e_{+}^{2} + e_{-}^{2} + e_{0}^{2})$, where the roots $e_{\pm}$ are complementary to $e_{0}$ (i.e., $e_{+} + e_{-} + e_{0} = 0$), we find
\begin{equation}
e_{+}^{2} \;+\; e_{-}^{2} \;=\; \frac{g_{2}}{2} \;-\; e_{0}^{2} \;=\;  \frac{20}{9}\,{\sf C}_{0}^{2} \;-\; 2\,{\sf C}_{0}^{2}\,\epsilon,
\label{eq:e_ac_1}
\end{equation}
while $g_{3} \equiv 4\,e_{+}e_{-}e_{0}$ yields
\begin{equation}
-\,2\,e_{+}e_{-} \;=\; -\,\frac{g_{3}}{2\,e_{0}} \;=\; \frac{3\,g_{3}}{4\,{\sf C}_{0}} \;=\; \frac{16}{9}\,{\sf C}_{0}^{2} \;-\; 2\,{\sf C}_{0}^{2}\,\epsilon.
\label{eq:e_ac_2}
\end{equation}
By combining Eqs.~\eqref{eq:e_ac_1}-\eqref{eq:e_ac_2}, we therefore find $(e_{+} - e_{-})^{2} = 4\,{\sf C}_{0}^{2}\,(1 - \epsilon)$, which, when using Eq.~\eqref{eq:e_c} and the constraint $e_{+} + e_{-} + e_{0} = 0$, can be solved as
\begin{equation}
e_{\pm} \;=\; \frac{1}{3}\,{\sf C}_{0} \;\pm\; \sqrt{{\sf C}_{0}^{2}\,(1 - \epsilon)}.
\label{eq:e_ac}
\end{equation}
We note that the amplitude $Q_{0} = \sqrt{p_{0}}$ is now expressed as $Q_{0}^{2} \equiv {\sf C}_{0} \pm \sqrt{{\sf C}_{0}^{2}\,(1 - \epsilon)} = e_{\pm} - e_{0}$ and the discriminant is
\begin{equation}
\Delta \;\equiv\; g_{2}^{3} - 27\,g_{3}^{2} \;=\; 64\,{\sf C}_{0}^{6}\,\epsilon^{2}\,(1 - \epsilon),
\label{eq:Delta}
\end{equation}
which is positive for $\epsilon < 1$. The selection of the cubic roots $(e_{1},e_{2},e_{3})$ from $(e_{+},e_{-},e_{0})$ will depend on the signs of $(g_{2},g_{3})$ according to Fig.~\ref{fig:roots_g2}. 

\subsection{Janda's Weierstrass elliptic solution}

We conclude this Appendix by summarizing Janda's work \cite{Janda_2018}, which contained a sign error that was corrected by Brizard \cite{Brizard_2019} (see Janda's response in Ref.~\cite{Janda_2019}). In his work, Janda \cite{Janda_2018} proposes the Weierstrass elliptic solution to the differential equation $\dot{q}^{2}  = \;=\; q^{4} - 2\,{\sf C}_{0}\,q^{2} + {\sf C}_{0}^{2}\epsilon$:
\begin{equation}
q(t) \;=\; q_{0} \left(1 \;+\; \frac{q_{1}}{\wp(t + t_{0}) + q_{2}}\right)
\label{eq:Janda_sol}
\end{equation} 
where $(q_{0},q_{1},q_{2})$ and $t_{0}$ are constants. We note that we have adapted Janda's notation $({\it c}_{0},{\it c}_{2})$ to match our notation: ${\it c}_{2} = {\sf C}_{0}$ and ${\it c}_{0} = \sqrt{{\sf C}_{0}^{2}(1 - \epsilon)}$, so that ${\it c}_{2}^{2} - {\it c}_{0}^{2} = {\sf C}_{0}^{2}\,\epsilon$. Because $q(t)$ must be the solution of the same ordinary differential equation, we substitute the expression $\dot{q}^{2} = q_{0}^{2}q_{1}^{2}\,\dot{\wp}^{2}/(\wp + q_{2})^{4}$ to obtain the Weierstrass differential equation \eqref{eq:wp_eq}, where
\begin{eqnarray}
q_{0} &=& \pm\sqrt{{\sf C}_{0} \;\pm\; \sqrt{{\sf C}_{0}^{2}\,(1 - \epsilon)}} \;=\; \pm\,Q_{0}^{\pm}, \label{eq:alpha} \\
q_{1} &=& q_{0}^{2} \;-\; {\sf C}_{0} \;=\; \sqrt{{\sf C}_{0}^{2}\,(1 - \epsilon)}, \label{eq:beta} \\
q_{2} &=& -\,\frac{1}{2}\,q_{0}^{2} \;+\; \frac{1}{6}\,{\sf C}_{0} \;=\; -\,\frac{1}{3}\,{\sf C}_{0} \;-\; \frac{1}{2}\,q_{1}, \label{eq:gamma}
\end{eqnarray}
which follow, respectively, from the requirements that the $\wp^{4}$-coefficient must vanish, the $\wp^{3}$-coefficient must be equal to 4 ($q_{1}$ is chosen to be positive), and the $\wp^{2}$-coefficient must also vanish. The new lattice invariants are now expressed as
\begin{eqnarray}
-\,\ov{g}_{2} &\equiv& 12\,q_{2}^{2} \;+\; 4\,(3\,q_{0}^{2} - {\sf C}_{0})\,q_{2} \;+\; 4\,q_{0}^{2}\,q_{1} \;=\; {\sf C}_{0}^{2}\,(1 - \epsilon) \;-\; \frac{4}{3}\,{\sf C}_{0}^{2}, \label{eq:g2_def} \\
-\,\ov{g}_{3} &\equiv& 4\,q_{2}^{3} \;+\;  2\,(3\,q_{0}^{2} - {\sf C}_{0})\,q_{2}^{2}  \;+\; 4\,q_{0}^{2}\,q_{1}\,q_{2} \;+\; q_{0}^{2}\,q_{1}^{2} \;=\; \frac{1}{27}\left[ 8\,{\sf C}_{0}^{3} \;-\; 9\,{\sf C}_{0}^{3}\,(1 - \epsilon)\right], \label{g3_def}
\end{eqnarray}
which yield the definitions
\begin{eqnarray}
\ov{g}_{2} &\equiv& \frac{1}{3}\,{\sf C}_{0}^{2}\,(1 + 3\,\epsilon) \;\equiv\; 2\,\left(\ov{e}_{1}^{2} \;+\; \ov{e}_{2}^{2} \;+\; \ov{e}_{3}^{2}\right), \label{eq:g2_e} \\
\ov{g}_{3} &\equiv& \frac{1}{27}\,{\sf C}_{0}^{3}\,(1 - 9\,\epsilon) \;\equiv\; 4\,\ov{e}_{1}\,\ov{e}_{2}\,\ov{e}_{3}, \label{eq:g3_e} \\
\ov{\Delta} &\equiv& \ov{g}_{2}^{3} \;-\; 27\,\ov{g}_{3}^{2} \;=\; {\sf C}_{0}^{6}\,\epsilon\,(1 - \epsilon)^{2} \;\equiv\; 16\,(\ov{e}_{1} - \ov{e}_{2})^{2}(\ov{e}_{2} - \ov{e}_{3})^{2}(\ov{e}_{3} - \ov{e}_{1})^{2}, \label{eq:delta_e}
\end{eqnarray}
which define the new cubic roots $(\ov{e}_{1},\ov{e}_{2},\ov{e}_{3})$ that are still required to satisfy the constraint $\ov{e}_{1} + \ov{e}_{2} + \ov{e}_{3} = 0$. We note that these expressions match Janda's results \cite{Janda_2018}: $\ov{g}_{2} = 
\frac{4}{3}\,{\it c}_{2}^{2} - {\it c}_{0}^{2}$, $\ov{g}_{3} = \frac{1}{3}\,{\it c}_{2} \left( {\it c}_{0}^{2} - \frac{8}{9}\,{\it c}_{2}^{2}\right)$, and $\ov{\Delta} = {\it c}_{0}^{4} \left({\it c}_{2}^{2} - {\it c}_{0}^{2}\right)$.

Comparing the parametrization \eqref{eq:g_2}-\eqref{eq:g_3} and \eqref{eq:Delta} with Eqs.~\eqref{eq:g2_e}-\eqref{eq:delta_e} shows that the Janda parametrization \eqref{eq:g2_e}-\eqref{eq:delta_e} will lead to different cubic roots $(\ov{e}_{1},\ov{e}_{2},\ov{e}_{3})$ and quarter-periods $(\ov{\omega}_{1},\ov{\omega}_{2},\ov{\omega}_{3})$. Here, the Weierstrass spectrum $(\ov{\omega}_{1},\ov{\omega}_{3})$, shown in Figs.~\ref{fig:omega_g2n}-\ref{fig:omega_13}, can be broken down into four regions for the orbit parameter $\epsilon$: (A) for $\epsilon < -\frac{1}{3}$, we find $(\ov{g}_{2} < 0, {\sf C}_{0}\ov{g}_{3} > 0, \ov{\Delta} < 0)$; (B) for $-\frac{1}{3} < \epsilon < 0$, we find $(\ov{g}_{2} > 0, {\sf C}_{0}\ov{g}_{3} > 0, \ov{\Delta} < 0)$; (C) for $0 < \epsilon < \frac{1}{9}$, we find $(\ov{g}_{2} > 0, {\sf C}_{0}\ov{g}_{3} > 0, \ov{\Delta} > 0)$; and (D) for $\epsilon > \frac{1}{9}$, we find $(\ov{g}_{2} > 0, {\sf C}_{0}\ov{g}_{3} > 0, \ov{\Delta} \geq 0)$, where $\ov{\Delta}$ vanishes at $\epsilon = 1$. Here, we use the notation ${\sf C}_{0}\ov{g}_{3} > 0$ to include the two cases $(\ov{g}_{3} > 0, {\sf C}_{0} > 0)$ and $(\ov{g}_{3} < 0, {\sf C}_{0} < 0)$.

Next, by inserting the generic cubic root $\ov{e}_{a} = {\sf C}_{0}/3$ into Eqs.~\eqref{eq:g2_e}-\eqref{eq:g3_e}, we obtain the generic complementary cubic roots $\ov{e}_{b,c} = -\,{\sf C}_{0}/6 \pm \sqrt{{\sf C}_{0}^{2}\,\epsilon/4}$, which implies that the cubic roots are real for all orbits with $\epsilon > 0$ (i.e., orbits I-III and V-VI), while only $\ov{e}_{a}$ remains real, with $\ov{e}_{b} = \ov{e}_{c}^{*}$, when $\epsilon < 0$ (i.e., orbits IV and VII). These root assignments agree with Janda's work \cite{Janda_2018}, where the cubic root $\ov{e}_{1} = {\sf C}_{0}/3 > 0$ is the only real root for all values of $\epsilon$, while $\ov{e}_{2,3} = -\,{\sf C}_{0}/6 \pm \sqrt{{\sf C}_{0}^{2}\,\epsilon/4}$ are real (or complex conjugate), for $\epsilon > 0$ (or $\epsilon < 0$). Here, we note that the selection of the cubic roots $(\ov{e}_{1},\ov{e}_{2},\ov{e}_{3})$ from $(\ov{e}_{a},\ov{e}_{b},\ov{e}_{c})$ will depend on the signs of $(\ov{g}_{2},\ov{g}_{3},\ov{\Delta})$ according to Fig.~\ref{fig:roots_g2}. 

Using the definitions \eqref{eq:alpha}-\eqref{eq:gamma}, Janda's Weierstrass solution \eqref{eq:Janda_sol} becomes
\begin{equation}
q(t) \;=\; Q_{0}^{\pm} \left[ 1 \;+\; \frac{q_{1}}{\wp(t + t_{0}) - \ov{e}_{a} - \frac{1}{2}\,q_{1}}\right] \;=\; Q_{0}^{\pm} \left[ \frac{\wp(t + t_{0}) - \ov{e}_{a} + \frac{1}{2}\,q_{1}}{\wp(t + t_{0}) - \ov{e}_{a} - \frac{1}{2}\,q_{1}}\right],
\label{eq:Janda_q}
\end{equation}
where $Q_{0}^{\pm} = \sqrt{{\sf C}_{0} \pm \sqrt{{\sf C}_{0}^{2}\,(1 - \epsilon)}} > 0$ is chosen as the positive root of Eq.~\eqref{eq:alpha}. By using the Weierstrass quarter-period identity \cite{NIST_Weierstrass}
\[ \wp\left(\frac{1}{2}\,\ov{\omega}_{a}\right) \;\equiv\; \ov{e}_{a} \;+\; \sqrt{(\ov{e}_{a} - \ov{e}_{b})\,(\ov{e}_{a} - \ov{e}_{c})} \;=\; \frac{1}{3}\,{\sf C}_{0} \;+\; \frac{1}{2}\,\sqrt{{\sf C}_{0}^{2}(1 - \epsilon)} \;=\; -\,q_{2}, \]
on the other hand, Janda's Weierstrass solution \eqref{eq:Janda_sol} can also be expressed as
\begin{equation}
q(t) \;=\; Q_{0}^{\pm} \left[1 \;+\; \frac{q_{1}}{\wp(t + t_{0}) - \wp\left(\frac{1}{2}\,\ov{\omega}_{a}\right)}\right] \;=\; Q_{0}^{\pm} \left[ \frac{\wp(t + t_{0}) - 2\,\ov{e}_{a} + \wp\left(\frac{1}{2}\,\ov{\omega}_{a}\right)}{\wp(t + t_{0}) - \wp\left(\frac{1}{2}\,\ov{\omega}_{a}\right)}\right].
\label{eq:q_Janda}
\end{equation}
For all unbounded orbits (III-IV,VII) with a single turning point, so that $q(0) = Q_{0}^{+}$ (i.e., $t_{0} = 0$), we obtain an immediate solution for the finite singularity time $T_{\infty} = \frac{1}{2}\,\ov{\omega}_{a}$. For the periodic orbit (II) with two turning points, we must select the two signs in Eq.~\eqref{eq:alpha}, with $Q_{0}^{-} = \sqrt{{\sf C}_{0} - \sqrt{{\sf C}_{0}^{2}\,(1 - \epsilon)}}$,  so that the piecewise-continuous solution
\[ q_{II}(t) \;=\; Q_{0}^{-} \left\{ \begin{array}{ll}
[\wp(t) - \ov{e}_{a} + \frac{1}{2}\,q_{1}]/[\wp(t) - \ov{e}_{a} - \frac{1}{2}\,q_{1}] & \hspace*{0.2in}(0 \leq t \leq \ov{\omega}_{a}) \\
 & \\
-\;[\wp(t - \ov{\omega}_{a}) - \ov{e}_{a} + \frac{1}{2}\,q_{1}]/[\wp(t) - \ov{e}_{a} - \frac{1}{2}\,q_{1}] &  \hspace*{0.2in}(\ov{\omega}_{a} \leq t \leq 2\,\ov{\omega}_{a}) 
\end{array} \right. \]
has a period of $2\,\ov{\omega}_{a}$. Here, $q_{II}(0) = Q_{0}^{-} = q_{II}(2\,\ov{\omega}_{a})$ and $q_{II}(\ov{\omega}_{a}) = -\,Q_{0}^{-}$. For unbounded orbits (I,V-VI) without a turning point, on the other hand, where we choose $q(0) = 0$, the initial-time constant $t_{0}$ is chosen in Eq.~\eqref{eq:q_Janda} from the expression
\begin{equation}
t_{0} \;\equiv\; \wp^{-1}\left[ 2\,\ov{e}_{a} - \wp\left(\frac{1}{2}\,\ov{\omega}_{a}\right)\right],
\label{eq:t_0}
\end{equation}
where $\wp^{-1}(z)$ denotes the inverse Weierstrass elliptic function, i.e., $\wp^{-1}(e_{k};g_{2},g_{3}) = \omega_{k}(g_{2},g_{3})$, and the finite singularity time occurs at \cite{Janda_2018,Janda_2019}
\begin{equation}
T_{\infty} = \frac{1}{2}\,\ov{\omega}_{a} \;-\; \wp^{-1}\left[ 2\,\ov{e}_{a} - \wp\left(\frac{1}{2}\,\ov{\omega}_{a}\right)\right].
\label{eq:T_inf0}
\end{equation}
Hence, all the orbits associated with solutions of the XMHD magnetic equation $\dot{q}^{2} \;=\; {\sf C}_{0}^{2}\epsilon - 2\,{\sf C}_{0}\,q^{2} + q^{4}$ can be represented in terms of Weierstrass elliptic solution \eqref{eq:q_Janda}.

Lastly, the Hall MHD solutions \eqref{eq:I_0}-\eqref{eq:V_0} involve exponentials of the integrated factor $\beta(t) = \int_{0}^{t}q(t')\,dt'$. We can now obtain an expression for the Hall MHD integrated factor $\beta(t) = \int_{0}^{t}q(t')\,dt'$, where
\begin{equation}
\beta(t) \;=\; Q_{0}^{\pm} \left[t \;+\; \int_{0}^{t}\frac{q_{1}\,dt'}{\wp(t' + t_{0}) - \wp\left(\frac{1}{2}\,\ov{\omega}_{a}\right)}\right].
\label{eq:beta_W}
\end{equation}
To evaluate the remaining integral, we use the Weierstrass integral identity \cite{Lawden}
\begin{equation}
\int_{t_{0}}^{t + t_{0}} \frac{du}{\wp(u) - \wp(v)} \;=\; \frac{1}{\wp^{\prime}(v)} \left[ 2\,\zeta(v)\;t \;+\; \ln\left(\frac{\sigma(t + t_{0} - v)}{\sigma(t + t_{0} + v)}\;\frac{\sigma(t_{0} + v)}{\sigma(t_{0} - v)}\right) \right],
\label{eq:W_int}
\end{equation}
where $v = \frac{1}{2}\,\ov{\omega}_{a}$ is a constant, while $\zeta(u) \equiv \sigma^{\prime}(u)/\sigma(u)$ and $\wp(u) \equiv -\,\zeta^{\prime}(u)$ are used to define the Weierstrass zeta and sigma functions \cite{NIST_Weierstrass,Lawden}. Since $\wp(\frac{1}{2}\,\ov{\omega}_{a}) = -\,q_{2}$, we use the definition $\wp^{\prime}(z) = \sqrt{4\,\wp^{3}(z) - g_{2}\,\wp(z) - g_{3}}$ to obtain
\[ \wp^{\prime}(\frac{1}{2}\,\ov{\omega}_{a}) \;=\; \sqrt{-\,4\,q_{2}^{3} + g_{2}q_{2} - g_{3}} \;=\; Q_{0}^{+}\,q_{1}. \]
The odd-parity zeta function is quasi-periodic: $\zeta(u + \omega_{k}) = \zeta(u) + \eta_{k}$, where $\eta_{k} \equiv \zeta(\omega_{k})$. Hence, $\zeta(\ov{\omega}_{a}/2) = \ov{\eta}_{a}/2$ and we find $2\,\zeta(\ov{\omega}_{a}/2)\;t 
= \ov{\eta}_{a}t$, and, thus, Eq.~\eqref{eq:W_int} becomes
\begin{equation}
\int_{t_{0}}^{t + t_{0}} \frac{du}{\wp(u) - \wp(\frac{1}{2}\,\ov{\omega}_{a})} \;=\; \frac{1}{Q_{0}^{+}\,q_{1}}\left[\ov{\eta}_{a}\,t \;+\; \ln\left(\frac{\sigma(T_{\infty} - t)}{\sigma(t + \ov{\omega}_{a} - T_{\infty})}\;
\frac{\sigma(\ov{\omega}_{a} - T_{\infty}))}{\sigma(T_{\infty})}\right) \right],
\label{eq:W_exp}
\end{equation}
where we substituted Eqs.~\eqref{eq:t_0}-\eqref{eq:T_inf0} and we used $\sigma(t - T_{\infty})/\sigma(-\,T_{\infty}) = \sigma(T_{\infty} - t)/\sigma(T_{\infty})$, which follows from the odd-parity of $\sigma(u)$. Hence, from Eq.~\eqref{eq:beta_W}, we obtain the final Weierstrass expression for the integrated factor (for unbounded orbits)
\begin{equation}
\beta(t) \;=\; (Q_{0}^{+} + \ov{\eta}_{a})\,t \;+\; \ln\left(\frac{\sigma(T_{\infty} - t)}{\sigma(t + \ov{\omega}_{a} - T_{\infty})}\;\frac{\sigma(\ov{\omega}_{a} - T_{\infty}))}{\sigma(T_{\infty})}\right) \;\equiv\; -\;\Gamma(t) \;+\; \ln\left(\frac{\sigma(T_{\infty} - t)}{\sigma(t + \ov{\omega}_{a} - T_{\infty})}\;\frac{\sigma(\ov{\omega}_{a} - T_{\infty}))}{\sigma(T_{\infty})}\right),
\label{eq:beta_W_final}
\end{equation}
where $\Gamma(t) = -\,(Q_{0}^{+} + \ov{\eta}_{a})\,t \equiv \gamma_{0}\,t$ can be eliminated by a suitable choice for $\gamma(t) = \gamma_{0}$ in our 2D self-similar model \eqref{eq:XMHD_coef}, so that the exponential factors $\exp[\pm\,2(\Gamma + \beta)]$ in Eqs.~\eqref{eq:I_gamma}-\eqref{eq:V_gamma} can be expressed solely as rational functions of the Weierstrass sigma function $\sigma(z)$. When $t \rightarrow T_{\infty} = \frac{1}{2}\,\ov{\omega}_{a} - t_{0}$, the expression \eqref{eq:beta_W_final} diverges (as expected), since $\lim_{t \rightarrow T_{\infty}}\;\sigma\left(T_{\infty} - t\right) = \infty$, while $\sigma(\ov{\omega}_{a})$ is finite. We can see that the calculation of the integrated factor \eqref{eq:beta_W_final} is, however, not as simple as the integrated factor involving the Jacobian elliptic functions as shown in Table \ref{tab:beta}, since it involves auxiliary Weierstrass elliptic functions $(\sigma,\zeta)$ as well as the inverse Weierstrass elliptic function $\wp^{-1}$. 

We now conclude this Appendix by noting that, while the Weierstrass elliptic representation of the orbit solutions shown in Figs.~\ref{fig:potential}-\ref{fig:portrait} is not unique, e.g., Eq.~\eqref{eq:q_P} versus Eq.~\eqref{eq:q_Janda}, the Jacobi elliptic representation shown in Table \ref{tab:Jac_orbit} is unique and leads to a simple set of Hall MHD solutions \eqref{eq:I_0}-\eqref{eq:V_0}, with $\exp(\pm\int_{0}^{t} b(t')\,dt')$ expressed in terms of simple rational functions of the Jacobi elliptic functions (Table \ref{tab:beta}).

\end{document}